# Scale-up of room-temperature constructive quantum interference from single molecules to self-assembled molecular-electronic films

Xintai Wang[a,e†], Troy L. R. Bennett[b,†], Ali Ismael[a,c,†], Luke A. Wilkinson[b], Joseph Hamill[d], Andrew J. P. White[b], Iain M. Grace[a], Tim Albrecht[d], Benjamin J. Robinson[a*], Nicholas J. Long[b*], Lesley F. Cohen[e*] and Colin J. Lambert[a*]

[a]Physics Department, Lancaster University, Lancaster, LA1 4YB, UK.
[b]Department of Chemistry, Imperial College London, MSRH, White City, London, W12 0BZ, UK.
[c]Department of Physics, College of Education for Pure Science, Tikrit University, Tikrit, Iraq.
[d]Department of Chemistry, Birmingham University, Edgbaston, Birmingham, B15 2TT, UK.
[e]The Blackett Laboratory, Imperial College London, South Kensington Campus, London, SW7 2AZ, UK.

[†]These authors contributed equally to this work
* e-mail: c.lambert@lancaster.ac.uk; l.cohen@imperial.ac.uk; n.long@imperial.ac.uk; b.j.robinson@lancaster.ac.uk

**Abstract:** The realization of self-assembled molecular-electronic films, whose room-temperature transport properties are controlled by quantum interference (QI), is an essential step in the scale-up QI effects from single molecules to parallel arrays of molecules. Recently, the effect of *destructive* QI (DQI) on the electrical conductance of self-assembled monolayers (SAMs) has been investigated. Here, through a combined experimental and theoretical investigation, we demonstrate chemical control of different forms of *constructive* QI (CQI) in cross-plane transport through SAMs and assess its influence on cross-plane thermoelectricity in SAMs. It is known that the electrical conductance of single molecules can be controlled in a deterministic manner, by chemically varying their connectivity to external electrodes. Here, by employing synthetic methodologies to vary the connectivity of terminal anchor groups around aromatic anthracene cores, and by forming SAMs of the resulting molecules, we clearly demonstrate that this signature of CQI can be translated into SAM-on-gold molecular films. We show that the conductance of vertical molecular junctions formed from anthracene-based molecules with two different connectivities differ by a factor of approximately 16, in agreement with theoretical predictions for their conductance ratio based on constructive QI effects within the core. We also demonstrate that for molecules with thiol anchor groups, the Seebeck coefficient of such films is connectivity dependent and with an appropriate choice of connectivity can be boosted by ~50%. This demonstration of QI and its influence on thermoelectricity in SAMs represents a critical step towards functional ultra-thin-film devices for future thermoelectric and molecular-scale electronics applications.

**Introduction:** Molecular electronic devices have the potential to deliver logic gates, sensors, memories and thermoelectric energy harvesters with ultra-low power requirements and sub-10 nm device footprints.[1-4] Single-molecule electronic junctions [5-12] and self-assembled monolayers[13-15] have been investigated intensively over the past few years, because their room-temperature electrical conductance has been shown to be controlled by destructive quantum interference (DQI).[16-20] More recently the effect of quantum interference on the Seebeck coefficient of single molecules has also been studied[21-26]. Figure 1 (A) illustrates an example of a room-temperature constructive quantum interference (CQI) effect, in which electrical current is injected into and collected from an anthracene molecular core, via the green arrows, or alternatively via the red arrows. Such a change in connectivity in a classical resistor network would lead to only a small change in electrical conductance. In contrast, theory predicts and experiment confirms[27-29] that the room temperature, single-molecule, low-bias electrical conductance $G_1$ for the green connectivity is approximately an order of magnitude greater than the conductance $G_2$ of the red connectivity. This is a clear signature of room-temperature phase-coherent transport and of the varying degrees of CQI for the two different connectivities (ESI-Fig. S28). The chemical realization of the green connectivity is molecule **1** of Fig. 1, in which the terminal groups attached to electrodes inject a current into the anthracene core via alkyne linkages. Similarly, molecule **2** is a realization of the red connectivity. **3** and **4** are alternative realisations of the red and green connectivities, in which the thioether terminal groups are replaced by thioacetate groups (which can be deprotected *in-situ* to grant terminal thiols for gold binding), which can further control interfacial coupling and energy level alignment between molecules and electrodes.[30,31] Our aim is to create self-assembled monolayers (SAMs) from these compounds, demonstrate that these single-molecule signatures of CQI can be translated into SAM-based devices and assess the effect of CQI on their Seebeck coefficients. We indeed find that the electrical conductances of SAMs formed from **1** and **3** are significantly higher than those of SAMs formed from **2** and **4**. We also measure and calculate the Seebeck coefficients of these SAMs and show that the sign and magnitude of their thermopower is determined by a combination of their connectivities and the nature of their (thioacetate or thioether) anchor groups.

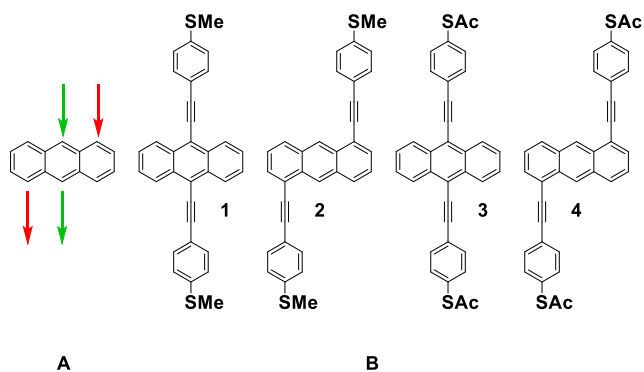

**Fig. 1 | Structures of studied molecules.** (**A**) A sketch of an anthracene core with connectivities 7,2' and 1,5'. (**B**) Chemical realisations of molecular wires with anthracene cores. **1** and **3** correspond to the 7,2' connectivity, while **2** and **4** correspond to the 1,5' connectivity.

**Results**

**Magic ratio theory**: Our choice of connectivities in Fig. 1 was guided by 'magic ratio theory,'[27] which predicts that the ratio $\frac{G_1}{G_2}$ of the low-bias, single-molecule conductances of **1** and **2** (**3** and **4**) should be $\frac{G_1}{G_2} = 16$ (ESI-Fig. S28). This simple theory illustrates how connectivity alone contributes to conductance ratios, without including chemical effects or Coulomb interactions. When the latter are included, recent studies[32] indicate that the qualitative trend in the ratio is preserved (*i.e.* that $\frac{G_1}{G_2} \gg 1$), but the precise value should be calculated using *ab initio* methods. Our aim is to determine if this single-molecule signature of QI is preserved or modified in a SAM, where intermolecular interactions are also expected to play a role.

Fig. 2 shows the frontier orbitals of **1** and **2**, and in agreement with magic ratio theory, confirms the presence of CQI, which occurs when the HOMO has different colours at the ends of the molecule (ie blue at one end and red at the other) and the LUMO has the same colour (ie red at both ends)[29,33-35]

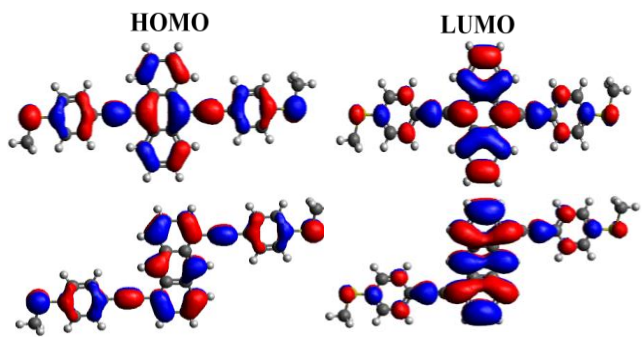

**Fig. 2 | Frontier orbitals for 1 and 2:** HOMO and LUMO orbitals for molecule 1 (top) and molecule 2 (bottom). (Orbitals for 3 and 4 are shown in the SI) Red (blue) corresponds to regions in space of positive (negative) orbital amplitude.

**Synthesis:** Molecules **1** and **2** bearing thioether termini could be synthesised from bromoanthracenes through the use of standard Sonagashira chemistry, however this same strategy could not be used to synthesise the thioacetate derivatives (**3** and **4**). This is due to a competing cyclo-oligermisation reaction that occurs when reacting a thioacetate-terminated phenylacetylide moiety in the presence of a palladium catalyst.[36] As a result of this, a trans-protection strategy was employed utilising a tert-butyl protected thiol. Initially, dibromoanthracenes were reacted with the alkyne of choice (either 4-ethynyl-tert-butylthioether or 4-ethynylthioanisole) to generate symmetrically disubstituted products (**1, 2, 3A** and **4A**). All compounds could be purified with flash column chromatography and were obtained in good yields (>60%). Thioacetate substituted anthracenes (**3** and **4**) were then obtained through *trans*-protection reactions of **3A** and **4A** respectively. Molecule **4** could be purified through the use of flash chromatography alone, however recrystallization was required to isolate molecule **3**, resulting in a slightly reduced yield. Further details can be found in the SI (section 1.3).

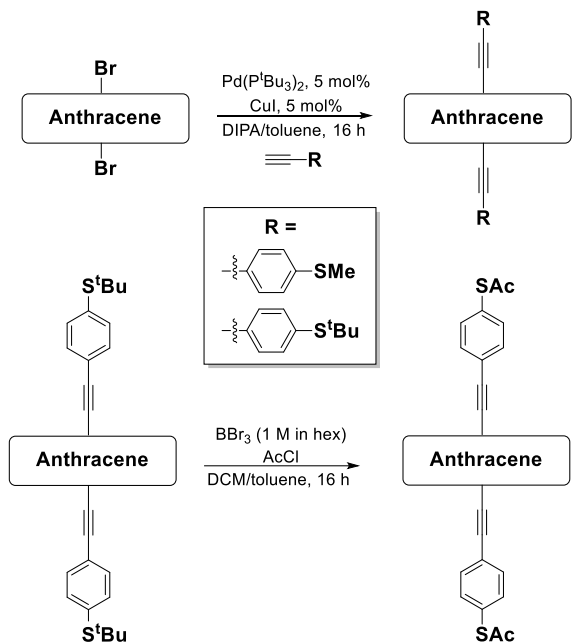

**Scheme. 1 | Synthesis of studied molecules.** A representative synthetic pathway illustrating the construction of symmetric anthracenes through the use of Sonagashira (top) and *trans*-protection (bottom) reactions.

**SAM formation and characterisation:** Deposited molecular films were characterized by atomic force microscopy (AFM), which suggested the formation of high-uniformity SAMs. All molecular films were grown on freshly prepared template stripped Au substrates[37,38] with a surface roughness of 80-150 pm (see Methods section). Averaged roughness, as measured across multiple random areas (ESI-Table S2), showed that SAMs of **1-4** conformally follow the underlying gold surface with comparable roughness. Film thicknesses were characterized by an AFM nano-scratching method[39-41] (details explained in the SI) with the thickness of all the films in the range of 1.1-1.4 nm; this thickness corresponds to a monolayer of molecules in a perpendicular configuration with a tilt angle of $30^0$-$50^0$. Larger area imaging of the sample surface further suggests that there were no multi-layered or un-covered regions (ESI-Table S4 and Figure S30); large scale uniformity was further confirmed through monitoring of film growth on a polished Au-coated quartz crystal microbalance (QCM). Here, a Sauerbrey analysis of the QCM frequency change indicates, that in all cases, the molecular occupation area corresponds to that expected for a single molecule in a closely packed SAM[42,43] (ESI-Table S4).

**Electrical and thermal Characterization:** Molecular conductance was characterized by conductive AFM (cAFM), where the number of molecules under the probe is estimated from the contact area between probe and sample surface (obtained via Hertz Model [44-46]) and the single-molecule occupation area obtained from QCM and AFM.

Conductance heat maps at low bias (-0.3 V to 0.3 V) for molecules **1-4** are shown in Figure 3 a and b, while Figure 3c shows the linear fit of thermal voltage vs. ΔT for different junction systems. The slope of the fit, $\frac{V_{Thermal}}{\Delta T}$, related with the Seebeck coefficient of the junction via equation: $S_{junction} = S_{probe-Au} - \frac{V_{Thermal}}{\Delta T}$ (the detailed number listed in Table 1). The two inset figures in Figure 3c show the thermal voltage distribution at differentΔT for a 1,5' anthracene junction with 2 thioacetate anchors and 2 thioether anchors, demonstrating that this exchange of anchor groups leads to a change in sign for the Seebeck coefficient.

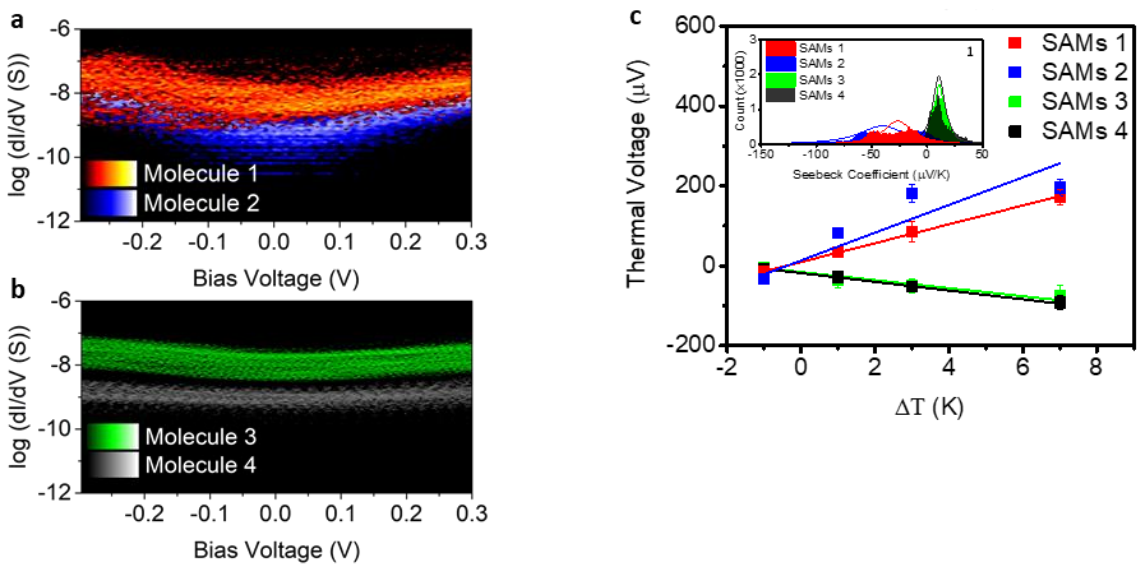

**Fig.3 | Electrical and thermoelectrical properties of SAMs**. (a,b) Heat map of molecular conductance (molecule 1, 2 (a), and 3, 4 (b)), bias voltage between -0.3 V to -0.3 V. (c) Linear fit plot of Thermal Voltage vs. ΔT (Tsample - Tprobe) for molecule 1-4. (inset Figure 1,) Seebeck coefficient distribution of SAMs 1-4.

**Table 1** | Experimental measurements and theoretical calculations ($E_F-E_F^{DFT}$ = –0.4 eV for (**1** and **2**), $E_F-E_F^{DFT}$ = +0.55 eV for (**3** and **4**), average, yellow-lines in Fig S28 in the ESI)

| M | Exp. (G/Go) | std | Theo. (G/Go) | Exp. S(μV/K) | std | Theo. S(μV/K) |
|---|---|---|---|---|---|---|
| 1 | 7.01E-5 | 9E-6 | 1.66E-4 | -21.6 | 6.0 | -20.0 |
| 2 | 6.88E-6 | 1E-6 | 1.05E-5 | -32.1 | 6.8 | -33.0 |
| 3 | 1.28E-4 | 5E-6 | 1.59E-4 | +11.0 | 0.4 | +12.5 |
| 4 | 9.0E-6 | 3E-6 | 1.00E-5 | +09.1 | 0.5 | +16.3 |

From the statistics of >200 different IV curves measured at different locations, the statistically-most-probable zero-bias differential conductance for molecule **1** is 10.2 times larger than that for molecule **2**, and 14.2 times larger than that of molecule **3**. These measured ratio are comparable with the value of 9 obtained from magic ratio theory and from the *ab initio* theoretical results, which we now present.

**Density functional theory:** To compute the electrical conductance of molecules **1-4**, we use density functional theory combined with the quantum transport code Gollum[47] to obtain the transmission coefficient $T(E)$ describing electrons of energy $E$ passing from the source to the drain electrodes, from which the room-temperature electrical conductance and Seebeck coefficient are determined, as described in the Theoretical Methods section.

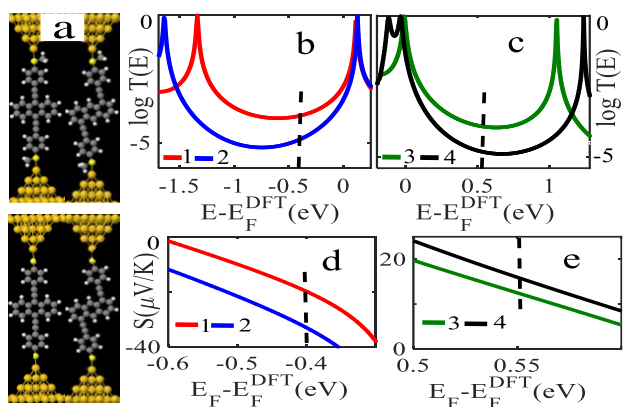

**Fig. 4 | Charge transport in molecular junctions.** (a) Schematic illustration of molecular junctions for **1**, **2**, **3** and **4**. (b and c) Transmission functions $T(E)$ for **1** (red solid-line), **2** (blue solid-line), **3** (green solid-line) and **4** (black solid-line). (d and e) Plots of the room-temperature Seebeck coefficients of **1-4** as a function of the Fermi energy $E_F$.

Fig. 4a shows that after structural relaxation, when placed between gold electrodes, the molecules adopt an angle corresponding to the measured tilt angle of the SAM (for different views see Fig. S29). Fig. 4b shows the computed transmission coefficients for all four junctions, while Figs. 4d-e show the corresponding Seebeck coefficients as a function of the Fermi energy $E_F$. Previous comparison between experiment and theory revealed that electron transport through polyaromatic hydrocarbons takes place near the middle of the energy gap between the highest occupied molecular orbital (HOMO) and the lowest unoccupied molecular orbital (LUMO),[27] and indeed we find that the closest agreement between theory and experiment is obtained for a Fermi energy near the mid-gap, indicated by the vertical dashed lines in Figs. 4b-c. The computed ratio of their transmission coefficients in gold-molecule-gold junctions (ESI-Fig. S28) for molecules 1 and 2 (similarly for 3 and 4) at $E = E_F^{Gold}$ is approximately 16. As described above, both molecules exhibit CQI near their gap centres and the conductance ratio arises from the different degrees of constructive QI associated with their different connectivities.[33,34,48-53] When the terminal groups of molecules are changed from thioethers to thioacetates, the transmission coefficients for molecules 3 and 4 show the same trend as those associated with molecules **1** and **2** (see Table 1).

In summary, through the rational design, synthesis and implementation of a new family of molecules, we have demonstrated that unequivocal signatures of single-molecule room-temperature CQI, contained in the connectivity-dependent conductance ratio of **1** and **2** (**3** and **4**), can be translated into self-assembled molecular films. With two different connectivities to the anthracene core, CQI effects lead to measured conductance ratios of $\left(G_1/G_2\right)_{Exp.} = 10.2, \left(G_3/G_3\right)_{Exp.} = 14.3,$ for SAMs formed from **1** compared to **2** (**3** and **4**), which is comparable with the magic ratio of 16 and the single-molecule DFT values of $\left(G_1/G_2\right)_{Theo.} = 15.8, \left(G_3/G_3\right)_{Theo.} = 16.0$. Furthermore, we show that the thermoelectrical performance of anthracene-based molecular films can be boosted by a judicious choice of connectivity to electrodes, combined with an optimal choice of terminal groups. Although the effect of CQI on the electrical conductance of SAMs was reported only recently[54], the above demonstration of CQI-controlled molecular films is the first report of CQI-boosted thermoelectricity. It opens the way to new design strategies for functional ultra-thin-film thermoelectric materials and electronic building blocks for future integrated circuits.

**Methods**

**Compound Synthesis and Characterization:** All reactions were performed with use of standard air-sensitive chemistry and Schlenk line techniques, under an atmosphere of nitrogen. No special precautions were taken to exclude air during any work-ups. All commercially available reagents were used as received from suppliers, without further purification. 4-Ethynylthioanisole, 4-(ethynyl)phenyl-tert-butylthioether and 1,5-dibromoanthracene were synthesised through adapted literature procedures.[10,29,54] Solvents used in reactions were collected from solvent towers sparged with nitrogen and dried with 3 Å molecular sieves, apart from DIPA, which was distilled on to activated 3 Å molecular sieves.

$^1$H and $^{13}$C{$^1$H} NMR spectra were recorded on a Bruker Avance 400 MHz spectrometer and referenced to the residual solvent peaks of either CDCl$_3$ at 7.26 and 77.16 ppm, respectively or DCM at 5.32 and 53.84. Coupling constants are measured in Hz. Mass spectrometry analyses were conducted by Dr. Lisa Haigh of the Mass Spectrometry Service, Imperial College London. Crystal structure analyses were conducted by Dr. Andrew White of the Crystallography Service, Imperial College London. Infrared spectra were recorded on a PerkinElmer Spectrum FT-IR spectrometer

**SAMs Fabrication:** For QCM: The QCM substrate (International Crystal Manufacturing, USA) was rinsed by acetone (>99%), methanol (>99%) and iso-propanol (>99%) in series and cleaned by oxygen plasma for 5 minutes. The stabilised, initial resonance frequency ($f_0$) of the cleaned QCM substrate was recorded. The cleaned QCM substrate was then immersed in 1 mM solution of molecules **1-4** in 1:2 ethanol:THF mixture (>99.9%, bubbling with nitrogen for 20 min to remove oxygen) from 12 hours to 48 hours. Optimised assembly times were established over multiple depositions. The substrate was subsequently rinsed by THF and ethanol several times to remove excess physisorbed molecules before drying in vacuum (10$^{-2}$ mbar, 40°C). The frequency of substrate after SAMs growth was again measured by the QCM. The equivalent measurement, where the QCM substrate was immersed in 1:2 ethanol:THF mixture without any molecules **1-4** present was also pre-formed as a reference.

**TS gold preparation for SPM:** A Si wafer (5 mm x 5 mm) was cleaned in an ultra-sonication bath with acetone, methanol and isopropanol in series, before cleaning with oxygen plasma for 5 minutes. The cleaned wafer was glued onto the top surface of a thermal evaporated gold sample previously grown on Si (100 nm thickness) with Epotek 353nd epoxy adhesive to form Si/Glue/Au/Si sandwich structure. The adhesive was cured for 40 minutes at 150 oC, then the original, bottom Si substrate was carefully removed using a sharp blade leaving an atomically-flat Au surface which was templated on the original Si surface. The prepared gold was scanned by AFM for 3-5 random spots for quality tests. For all cases, only the substrates with roughness below 0.2 nm were used for SAMs growth.

**SAMs Growth:** Following the optimised procedure for QCM, the gold was immersed in solution immediately after cleavage without any further treatment for 12 h (molecules 1, 2 and 4) and 24 h (molecule 3). The substrates were rinsed after molecular assembly by ThF and ethanol and dried in vacuum for 12 hours (10-2 mbar, 40oC).

**SAMs characterization:** SAM topography was characterized by AFM (MultiMode 8, Bruker Nanoscience) in peak force mode, a low force intermittent-contact mode with combines high resolution imaging, sample nanomechanical information and low sample damage. The peak force setpoint was set to the range of 500 pN to 1 nN and the scan rate was set to 1 Hz. The nano-scratching was performed in contact mode at high set force (F = 15 - 40 nN) using a soft probe (Multi-75-G, k = 3 N/m) to 'sweep away' the molecular film from a defined area ( A = 300 nm x 300 nm). The topography of sample after scratching was again characterized in peak force mode, the scratched window is easily observed. Nano-scratching was also conducted on a bare gold sample under the same conditions to ensure no gold is scratched away in used force range. The height difference between the scratched part and un-scratched part indicates the thickness of SAMs.

**Conductive AFM (cAFM):** The electrical transport properties of the SAMs were characterized by a custom cAFM system. The cAFM setup is based on a multi-mode8 AFM system (Bruker nanoscience). The bottom gold substrate was used as the source, and a Pt/Cr coated probe (Multi75 E, BugetSensor) was used as the drain. The force between probe and molecule was controlled at 2 nN, as this force is strong enough for the probe to penetrate through the water layer on the sample surface but not too strong to destroy the molecular thin film. The driven bias was added between the source and drain by a voltage generator (Aglient 33500B), the source to drain current was amplified by a current pre-amplifier (SR570, Stanford Research Systems), and the IV characteristics of the sample was collected by the computer.

**Thermal-Electrical Atomic Force Microscopy (ThEFM):** The Seebeck coefficients of SAMs were obtained by a ThEFM modified from the cAFM system used for electrical transport measurement. A peltier stage driven by a voltage generator (Aglient 33500B, voltage amplified by a wide band amplifier) was used to heat up and cool, thus a temperature difference can be created between sample and probe. The sample temperature was measured by a Type T thermal couple, and the probe temperature was calibrated by using an SThM (scanning thermal microscopy) probe (KNT SThM 2an) under the same conditions (F = 2nN). We made an assumption that the SThM probe and the cAFM probe have similar probe temperatures at the apex part when finding contact with the molecules. The thermal voltage between sample and probe was amplified by high impedance differential pre-amplifier (SR551, Stanford Research Systems), and recorded by a computer.

**Computational details:** The ground state Hamiltonian and optimized geometry of each molecule was obtained using the density functional theory (DFT) code.[55] The local density approximation (LDA) exchange correlation functional was used along with double zeta polarized (DZP) basis sets and the norm conserving pseudo potentials. The real space grid was defined by a plane wave cut-off of 250 Ry. The geometry optimization was carried out to a force tolerance of 0.01 eV/Å. This process was repeated for a unit cell with the molecule between gold electrodes where the optimized distance between Au and the pyridine anchor group was found to be 2.3 Å, whereas Au and SMe 2.7 Å. From the ground state Hamiltonian, the transmission coefficient, the room temperature electrical conductance $G$ and Seebeck coefficient $S$ was obtained, as described in section 2.4, 2.7 and 2.8 in the ESI. We model the properties of a single molecule in the junction as pervious works[56] have shown the calculated conductance of a SAM differs only slightly from that of single molecules

## ACKNOWLEDGMENTS

Support from the UK EPSRC is acknowledged, through grant nos. EP/N017188/1, EP/M014452/1, EP/P027156/1 and EP/N03337X/1. Support from the European Commission is provided by the FET Open project 767187 – QuIET. A.I is grateful for financial assistance from Tikrit University (Iraq), and the Iraqi Ministry of Higher Education (SL-20). NL is grateful for a Royal Society Wolfson Research Merit Award. LC and X.W acknowledge FSRF funding.

## AUTHOR INFORMATION

**Correspondence and requests for materials** should be addressed to C.L., L.C., N.L. and B.R.

**Author Contributions**

C.L., L.C., N.L., T.A. and B.R. conceived and designed the experiments. X.W. performed most of the device fabrication, characterization and data analysis. A.I. performed the theoretical simulation and data analysis. L.W. and T.B. synthesized the molecules. A.W. performed X-ray crystallography measurements and structural refinement. X.W. performed the AFM studies. C.L, L.C., N.L., T.A. and B.R. supervised the research. C.L, L.C., N.L., and B.R. co-wrote the paper. All authors discussed the results and commented on the manuscript.

# Supplementary Information

# Scale-up of room-temperature constructive quantum interference from single molecules to self-assembled molecular-electronic films

Xintai Wang[a,†], Troy L. R. Bennett[b,†], Ali Ismael[a,c,†], Luke A. Wilkinson[b], Joseph Hamil[d], Andrew J. P. White[b], Tim Albrecht[d], Benjamin J. Robinson[a*], Nicholas J. Long[b*], Lesley F. Cohen[e*] and  Colin J. Lambert[a*]

[a]Physics Department, Lancaster University, Lancaster, LA1 4YB, UK.

[b]Department of Chemistry, Imperial College London, MSRH, White City, London, W12 0BZ, UK.

[c]Department of Physics, College of Education for Pure Science, Tikrit University, Tikrit, Iraq.

[d]Department of Chemistry, Birmingham University, Edgbaston, Birmingham, B15 2TT, UK.

[e]The Blackett Laboratory, Imperial College London, South Kensington Campus, London, SW7 2AZ, UK.

†These authors contributed equally to this work

*To whom correspondence should be addressed. e-mail: c.lambert@lancaster.ac.uk; l.cohen@imperial.ac.uk; n.long@imperial.ac.uk; b.j.robinson@lancaster.ac.uk

## 1. Experimental

### 1.1 Materials and Methods

All reactions were performed with the use of standard air-sensitive chemistry and Schlenk line techniques, under an atmosphere of nitrogen. No special precautions were taken to exclude air during any work-ups. All commercially available reagents were used as received from suppliers, without further purification. 4-Ethynylthioanisole, 4-(ethynyl)phenyl-tert-butylthioether and 1,5-dibromoanthracene were synthesised through adapted literature procedures.[1-3] Solvents used in reactions were collected from solvent towers sparged with nitrogen and dried with 3 Å molecular sieves, apart from DIPA, which was distilled onto activated 3 Å molecular sieves under nitrogen.

### 1.2 Instrumentation

$^1H$ and $^{13}C\{^1H\}$ NMR spectra were recorded with a Bruker Avance 400 MHz spectrometer and referenced to the residual solvent peaks of $CDCl_3$ at 7.26 and 77.16 ppm, respectively. Coupling constants are measured in Hz. Mass spectrometry analyses were conducted by Dr. Lisa Haigh of the Mass Spectrometry Service, Imperial College London. Crystal structure analyses were conducted by Dr. Andrew White of the Crystallography Service, Imperial College London. Infrared spectra were recorded on a PerkinElmer Spectrum FT-IR spectrometer.

### 1.3 Synthesis

**General procedure for the coupling of terminal alkynes to bromoanthracenes –** specific details regarding molar equivalents and column conditions are reported below.

This synthetic procedure is adapted from a published method for Sonogashira coupling.[4] A Schlenk tube was charged with dibromoanthracene, terminal alkyne, CuI (5 mol%) and Pd(P-$^t$Bu$_3$)$_2$ (5 mol%) then placed under an inert atmosphere. Anhydrous DIPA and toluene were added to the reaction vessel via cannula and the reaction was stirred overnight at room temperature to generate a bright orange/yellow precipitate. Removal of the solvent *in vacuo* led to a dark brown crude material which can be purified by column chromatography.

**9,10-Di(4-(ethynyl)thioanisole)anthracene (1)[5]**

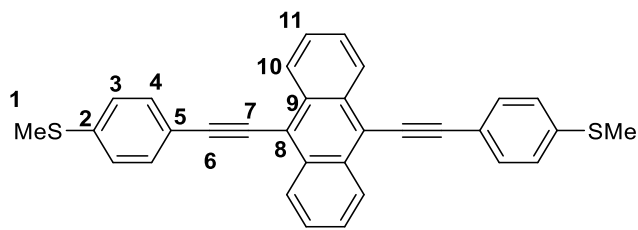

4-(Ethynyl)thioanisole (0.20 g, 1.35 mmol), 9,10-dibromoanthracene (1.81 g, 5.40 mmol), CuI (0.01 g, 0.07 mmol) and Pd(P-$^t$Bu$_3$)$_2$ (0.03 g, 0.07 mmol) gave an orange-brown solid which was purified by chromatography on a silica column, eluted with n-hexane/THF (1:0 → 1:1 v/v) to give the product as an orange solid (0.20 g, 0.85 mmol, 63%).

**$^1$H NMR** (CDCl$_3$, 298 K, 400 MHz): δ$_H$ = 8.69 (dd, $^3J_{HH}$ = 6.8, $^4J_{HH}$ = 3.2 Hz, 4H, *H10*), 7.70 (d, $^3J_{HH}$ = 8.4 Hz, 4H, *H4*), 7.67 (dd, $^3J_{HH}$ = 6.8, $^4J_{HH}$ = 3.2 Hz, 4H, *H11*), 7.32 (d, $^3J_{HH}$ = 8.4 Hz, 4H, *H3*), 2.55 (s, 6H, *H1*) ppm; **$^{13}$C {$^1$H} NMR** (CDCl$_3$, 298 K, 100 MHz): δ$_C$ = 140.2 (Ar-C-C), 132.2 (Ar-C-C), 132.1 (Ar-C-H), 127.4 (Ar-C-H), 126.9 (Ar-C-H), 126.2 (Ar-C-H), 119.8 (Ar-C-C), 118.6 (Ar-C-C), 102.5 (-C≡C-), 86.9 (-C≡C-), 15.6 (S-CH$_3$) ppm; **IR**: 2195 (-C≡C-) cm$^{-1}$; **MS** ES+: calcd. for C$_{32}$H$_{22}$S$_2$ [M]+ 469.1079; found. 469.1077.

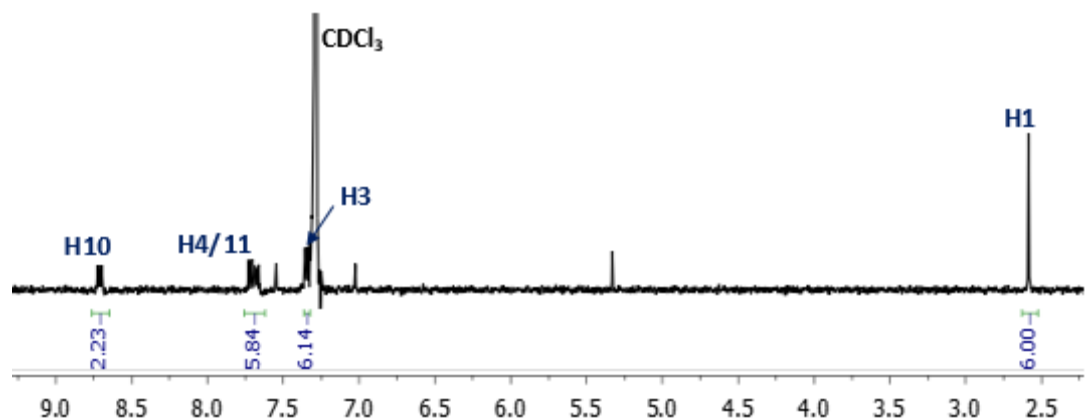

**Fig**ure **S1**: The $^1$H NMR spectrum of **1** in CDCl$_3$

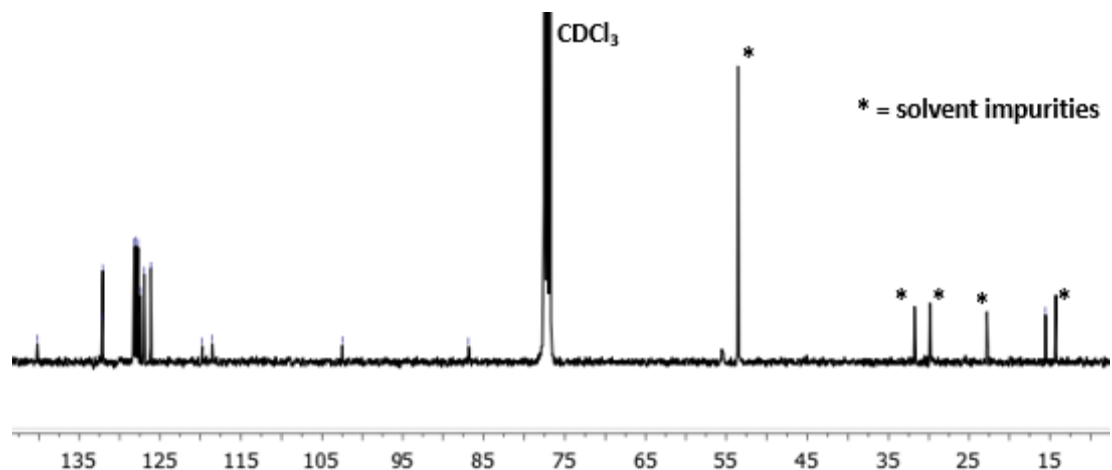

Figure S2: The ¹³C{¹H} NMR spectrum of 1 in CDCl₃

**1,5-Di(4-(ethynyl)thioanisole)anthracene (2)**

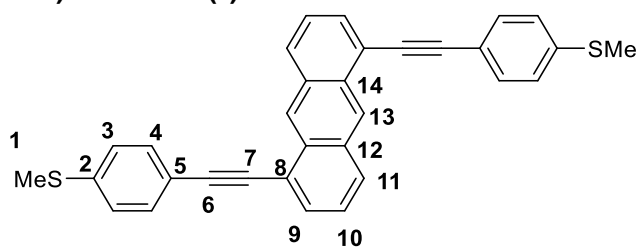

4-(Ethynyl)thioanisole (0.15 g, 1.01 mmol), 1,5-dibromoanthracene (0.10 g, 0.30 mmol), CuI (0.01 g, 0.03 mmol) and Pd(P-ᵗBu₃)₂ (0.02 g, 0.03 mmol) gave an orange-brown solid which was purified by chromatography on a silica column, eluted with n-hexane/THF (1:0 → 1:1 v/v) to give the product as an orange solid (0.11 g, 0.24 mmol, 79%).

**¹H NMR** (CDCl₃, 298 K, 400 MHz): $\delta_H$ = 8.99 (s, 2H, *H13*), 8.10 (dd, $^3J_{HH}$ = 4.4, $^4J_{HH}$ = 1.6 Hz, 2H, *H9*), 7.80 (dd, $^3J_{HH}$ = 6.8, $^4J_{HH}$ = 1.2 Hz, 2H, *H11*), 7.63 (d, $^3J_{HH}$ = 8.0 Hz, 4H, *H4*), 7.55-7.45 (m, 2H, *H10*), 7.30 (d, $^3J_{HH}$ = 8.0 Hz, 4H, *H3*), 2.55 (s, 6H, *H1*) ppm; **¹³C{¹H} NMR** (CDCl₃, 298 K, 100 MHz): $\delta_C$ = 139.8 (Ar-C-C), 132.1 (Ar-C-H), 131.4 (Ar-C-C), 130.7 (Ar-C-H), 129.6 (Ar-C-H), 126.1 (Ar-C-H), 125.9 (Ar-C-H), 125.3 (Ar-C-H), 121.1 (Ar-C-C), 119.8 (Ar-C-C), 94.9 (-C≡C-), 87.9 (-C≡C-), 15.6 (S-CH₃) ppm; **IR**: 2207 (-C≡C-) cm⁻¹; **MS** ES+: calcd. for C₃₂H₂₂S₂ [M]+ 469.1074; found. 469.1081.

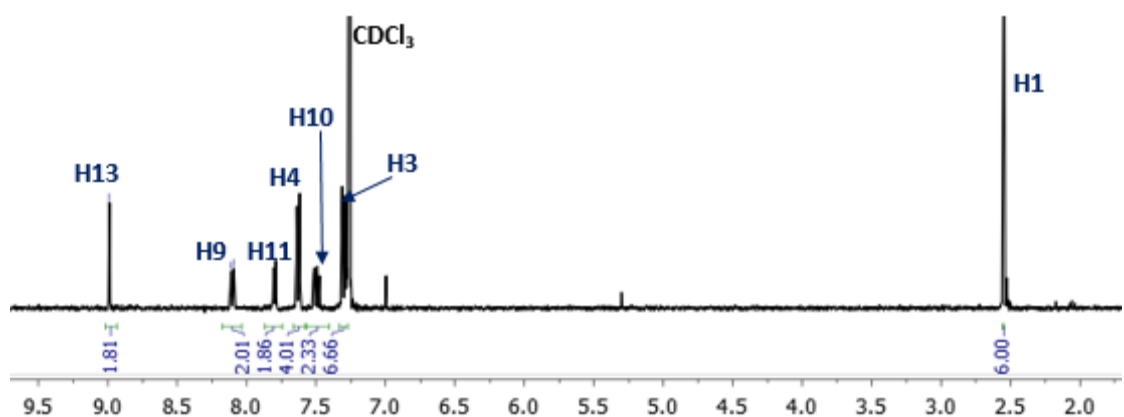

Figure S3: The ¹H NMR spectrum of 2 in CDCl₃

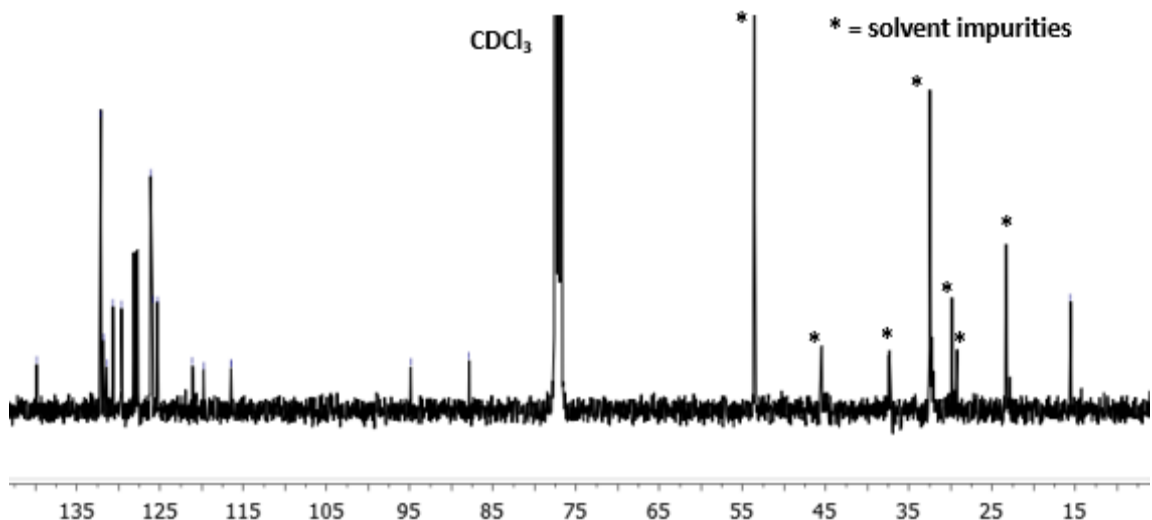

**Figure S4**: The $^{13}C\{^1H\}$ NMR spectrum of **2** in CDCl$_3$

**9,10-Di(4-(ethynyl)phenyl-*tert*-butylthioether)anthracene (3A)**[2]

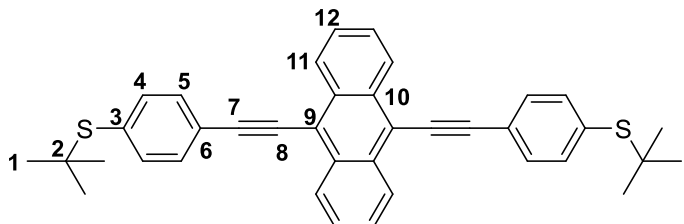

9,10-Dibromoanthracene (0.15 g, 0.45 mmol), 4-(ethynyl)phenyl-*tert*-butylthioether (0.21 g, 1.12 mmol), CuI (0.01 g, 0.05 mmol) and Pd(P-$^t$Bu$_3$)$_2$ (0.02 g, 0.05 mmol) gave an orange-brown solid which was purified by chromatography on a silica column, eluting with n-hexane/DCM (1:0→8:2) to give the product as an orange solid (0.19 g, 0.34 mmol, 76%).

**$^1$H NMR** (CDCl$_3$, 298 K, 400 MHz): $\delta_H$ = 8.72-8.67 (m, 4H, *H*11), 7.74 (d, $^3J_{HH}$ = 8.4 Hz, 4H, *H*5), 7.68-7.63 (m, 4H, *H*12), 7.63 (d, $^3J_{HH}$ = 8.4 Hz, 4H, *H*4), 1.35 (s, 18H, *H*1) ppm; **$^{13}$C{$^1$H} NMR** (CDCl$_3$, 298 K, 100 MHz): $\delta_C$ = 137.6 (Ar-C-H), 134.0 (Ar-C-C), 132.3 (Ar-C-C), 131.7 (Ar-C-H), 127.4 (Ar-C-H), 127.1 (Ar-C-H), 123.9 (Ar-C-C), 118.6 (Ar-C-C), 102.1 (-C≡C-), 88.1 (-C≡C-), 46.8 (S-C-C), 31.2 (CH$_3$) ppm; **IR**: 2194 (-C≡C-) cm$^{-1}$; **MS** APCI: calcd. [M]+555.2175; found. 555.2171.

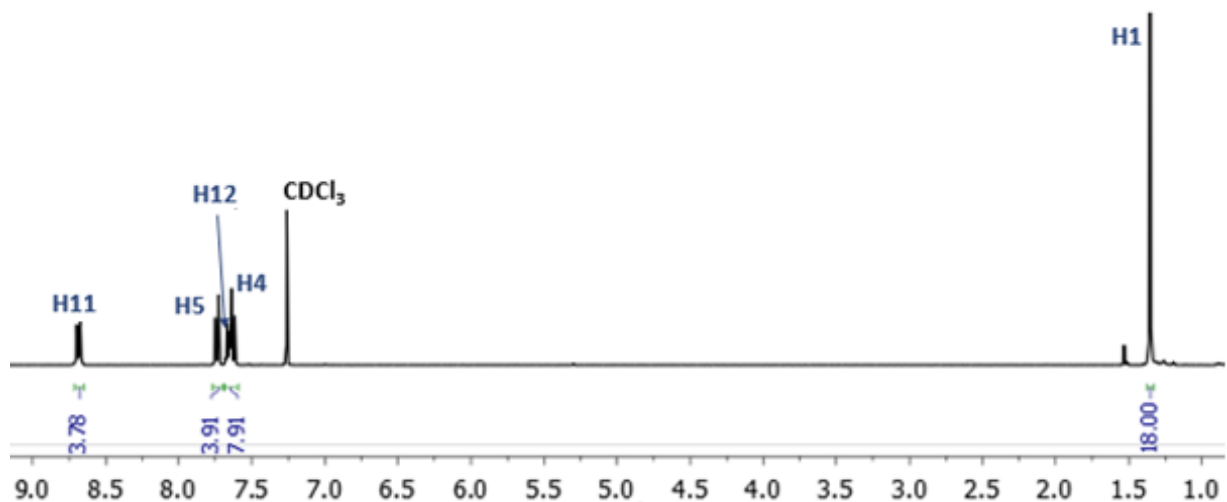

Figure S5: The ¹H NMR spectrum of 3A in CDCl₃

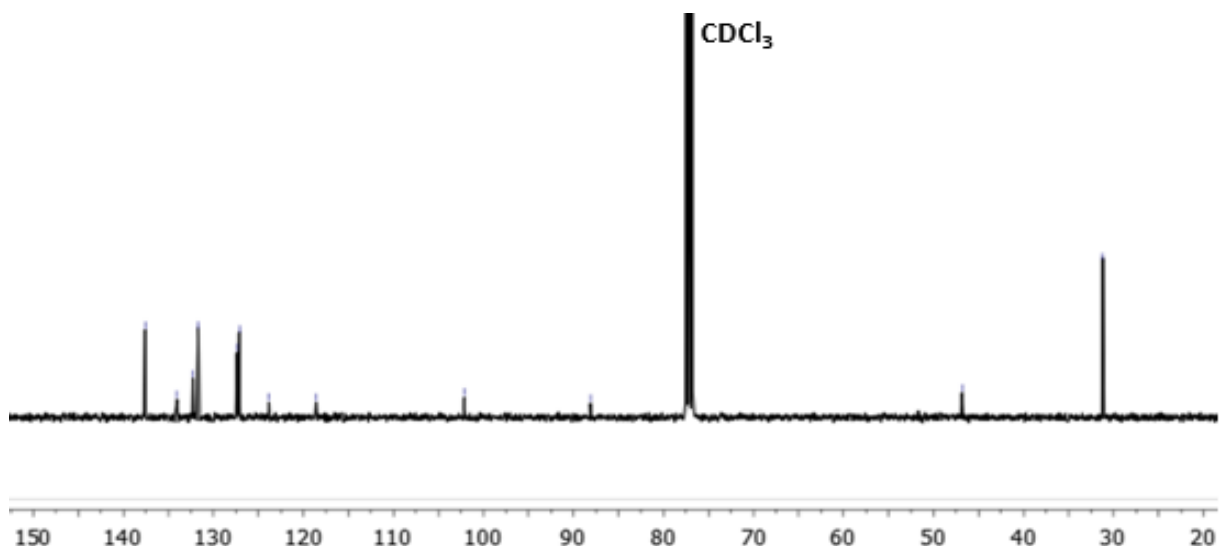

Figure S6: The ¹³C{¹H} NMR spectrum of 3A in CDCl₃

**9,10- Di(4-(ethynyl)phenylthioacetate)anthracene (3)[2]**

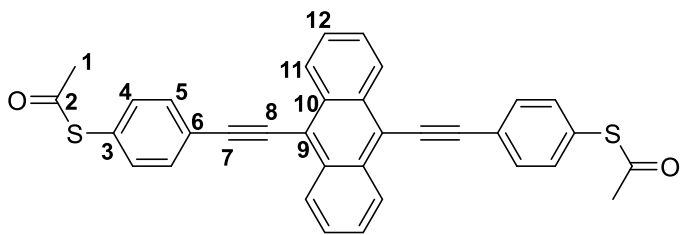

Synthesised according to an adapted literature procedure.[2] (**3A**) (0.08 g, 0.14 mmol) was dissolved in DCM (30 mL) and toluene (30 mL). Acetyl chloride (1 mL) was added and the solution was degassed for 20 minutes. BBr₃ (1 M in hexanes, 0.72 mL, 0.72 mmol) was added and the solution was stirred overnight at room temperature. The solvent was removed *in vacuo* and the crude product was exposed to chromatography on a silica column, eluting with n-hexane/DCM (1:0 → 1:1). The product was washed with hexane (3 x 100 mL) and recrystallized from DCM to yield dark orange crystals (0.03 g, 0.06 mmol, 43%).

**¹H NMR** (CDCl₃, 298 K, 400 MHz) $\delta_H$ = 8.71-8.64 (m, 4H, *H*11), 7.81 (d, $^3J_{HH}$ = 8.4 Hz, 4H, *H*5), 7.69-7.62 (m, 4H, *H*12), 7.51 (d, $^3J_{HH}$ = 8.4 Hz, 4H, *H*4), 2.48 (s, 6H, *H*1) ppm; **¹³C {¹H} NMR** (CDCl₃, 298 K, 100 MHz): $\delta_C$ = 193.4 (-C=O), 134.4 (Ar-C-H), 132.2 (Ar-C-H), 132.2 (Ar-C-C), 128.6 (Ar-C-C), 127.2 (Ar-C-H), 127.0 (Ar-C-H), 124.6 (Ar-C-C), 118.4 (Ar-C-C), 101.7 (-C≡C-), 88.1 (-C≡C-), 30.4 (CH₃) ppm; **IR**: 2195 (-C≡C-), 1692 (-C=O) cm⁻¹; **MS** ES+: calcd. [M]+ 527.1134; found. 527.1128.

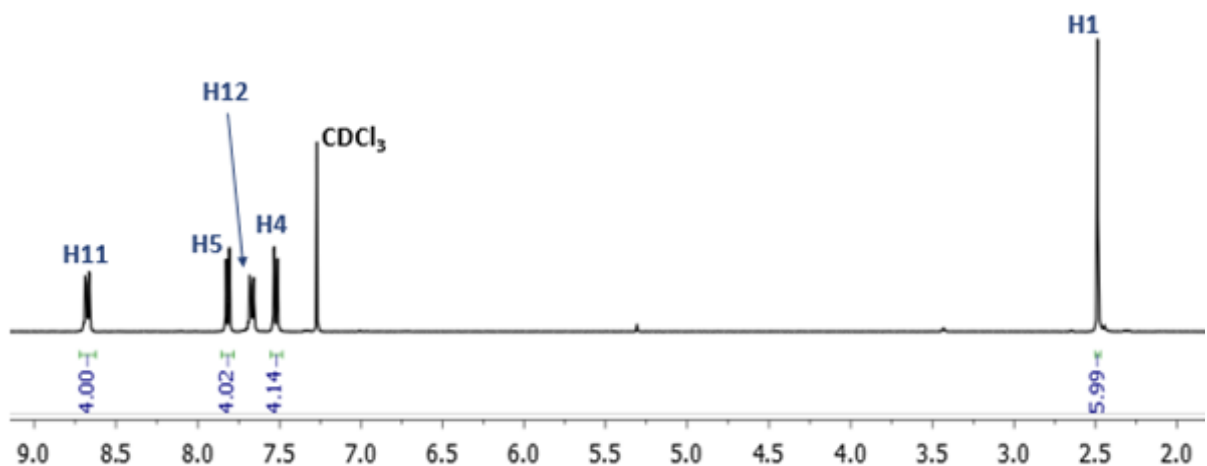

**Fig**ure S7: The ¹H NMR spectrum of **3B** in CDCl₃

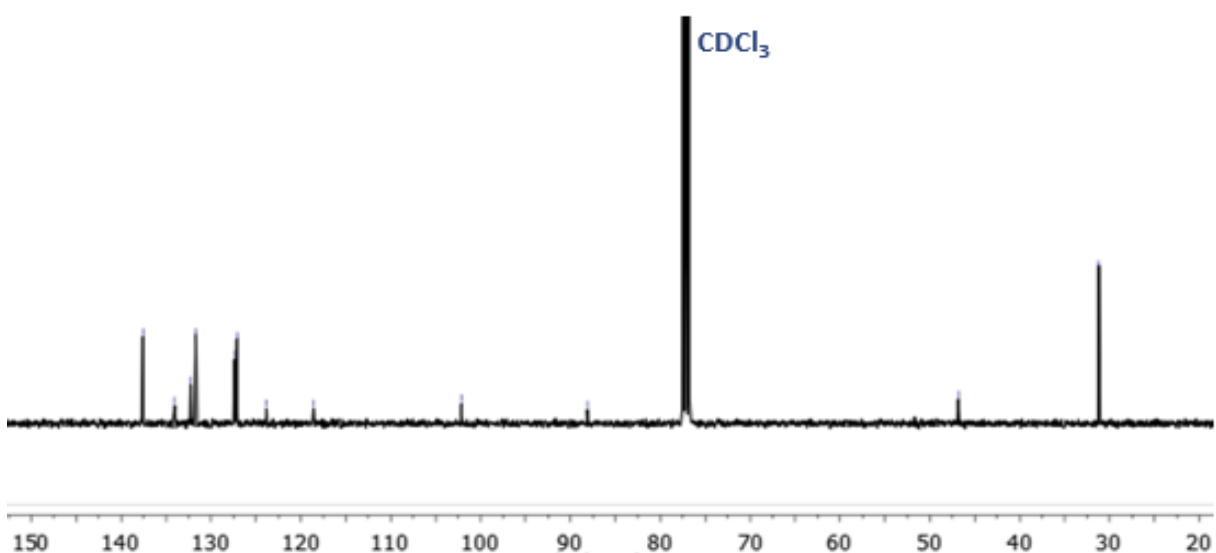

**Fig**ure S8: The ¹³C{¹H} NMR spectrum of **3B** in CDCl₃

**1,5- Di(4-(ethynyl)phenyl-*tert*-butylthioether)anthracene (4A)**

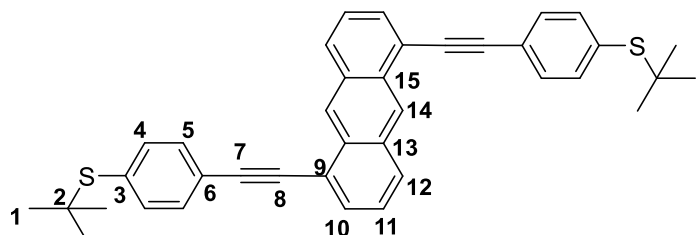

Synthesised according to an adapted literature procedure.[2] 1,5-Dibromoanthracene (0.15 g, 0.45 mmol), 4-(ethynyl)phenyl-*tert*-butylthioether (0.21 g, 1.12 mmol), CuI and Pd(P-ᵗBu₃)₂ (0.02 g, 0.05 mmol) gave a yellow-brown solid which was purified by chromatography on a silica column, eluting with n-hexane/DCM (1:0 → 4:1) to give the product as a yellow solid (0.22 g, 0.39 mmol, 87%).

**¹H NMR** (CDCl₃, 298 K, 400 MHz): $\delta_H$ = 9.00 (s, 2H, *H*14), 8.13 (dd, $^3J_{HH}$ = 4.4, $^4J_{HH}$ = 1.6 Hz, 4H, *H*10), 7.82 (dd, $^3J_{HH}$ = 6.8, $^4J_{HH}$ = 1.2 Hz, 2H, *H*12), 7.68 (d, $^3J_{HH}$ = 8.4 Hz, 4H, *H*5) 7.61 (d, $^3J_{HH}$ = 8.4 Hz, 4H, *H*4), 7.53-7.48 (m, 2H, *H*11), 1.34 (s, 18H, *H*1) ppm; **¹³C {¹H} NMR** (CDCl₃, 298 K, 100 MHz): $\delta_C$ = 137.5 (Ar-C-H), 133.7

(Ar-C-C), 131.8 (Ar-C-H), 131.8 (Ar-C-H), 131.4 (Ar-C-C), 130.9 (Ar-C-H), 129.9 (Ar-C-C), 125.9 (Ar-C-C), 125.4 (Ar-C-H), 123.9 (Ar-C-C), 120.9 (Ar-C-C), 94.5 (-C≡C-), 89.3 (-C≡C-), 46.8 (S-C-), 31.2 (CH$_3$) ppm; **IR**: 2202 (-C≡C-) cm$^{-1}$; **MS** APCI: calcd. [M]+555.2175; found. 555.2173.

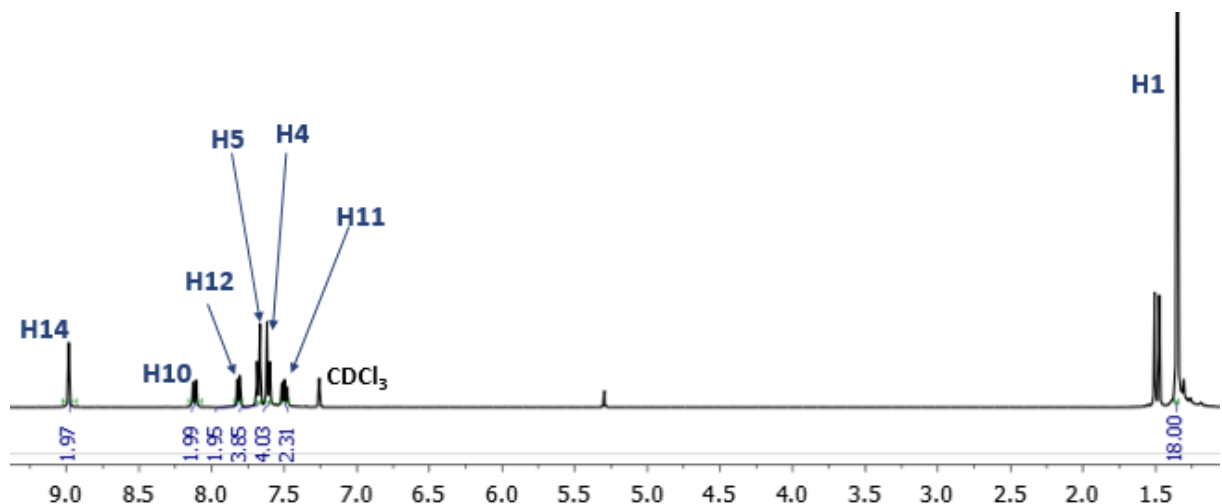

**Fig**ure **S9**: The $^1$H NMR spectrum of **4A** in CDCl$_3$

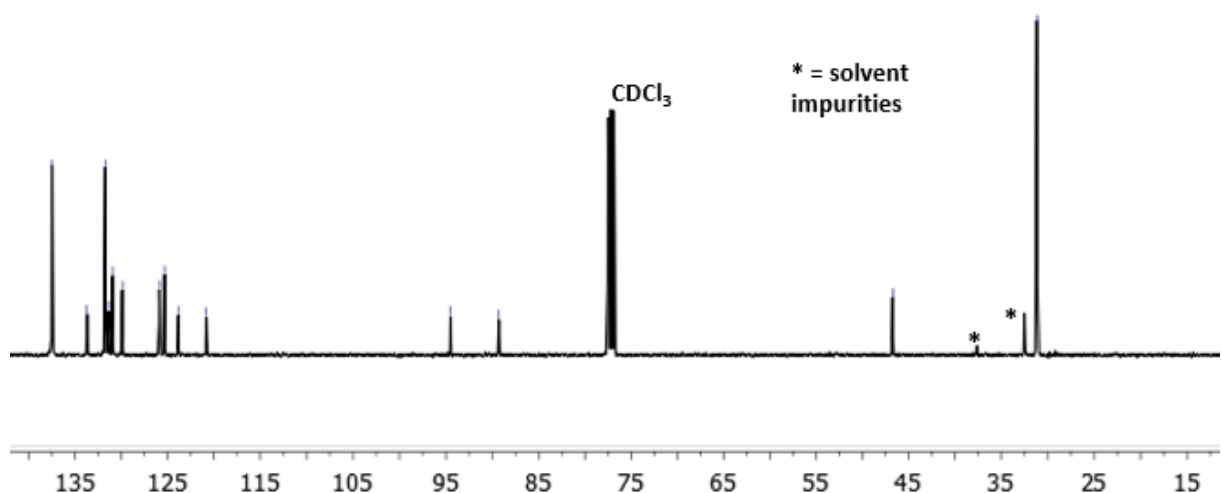

**Fig**ure **S10**: The $^{13}$C{$^1$H} NMR spectrum of **4A** in CDCl$_3$

**1,5- Di(4-(ethynyl)phenylthioacetate)anthracene (4)**

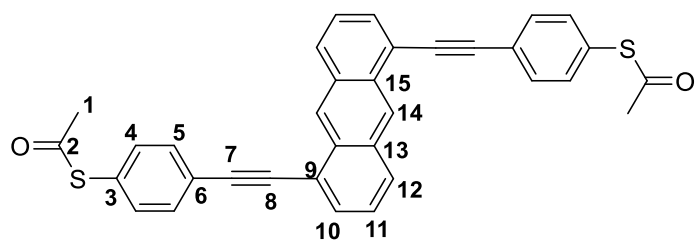

Synthesised according to an adapted literature procedure.[2] (**4A**) (0.08 g, 0.14 mmol) was dissolved in DCM (30 mL) and toluene (30 mL). Acetyl chloride (1 mL) was added and the solution was degassed for 20 minutes. BBr$_3$ (1 M in hexanes, 0.72 mL, 0.72 mmol) was added and the solution was stirred overnight at room temperature. The solvent was removed *in vacuo* and the crude product was exposed to chromatography on a silica column, eluting with chloroform to give the product as a yellow solid (0.05 g, 0.09 mmol, 63%).

**$^1$H NMR** (CDCl$_3$, 298 K, 400 MHz): δ$_H$ = 8.98 (s, 2H, *H*14), 8.13 (d, $^3J_{HH}$ = 4.4 Hz, 2H, *H*10), 7.81 (d, $^3J_{HH}$ = 4.4 Hz, 2H, *H*12), 7.75 (d, $^3J_{HH}$ = 8.4 Hz, 2H, *H*5), 7.54-7.49 (m, 6H, *H*4/11), 7.49 (d, $^3J_{HH}$ = 8.4 Hz, 2H), 2.47 (s, 6H) ppm; **$^{13}$C{$^1$H} NMR** (CDCl$_3$, 298 K, 100 MHz): δ$_C$ = 193.6 (-C=O), 134.5 (Ar-C-H), 132.5 (Ar-C-H), 131.8 (Ar-C-C), 131.4 (Ar-C-H), 131.1 (Ar-C-C), 130.1 (Ar-C-H), 128.5 (Ar-C-C), 125.9 (Ar-C-H), 125.4 (Ar-C-H), 124.8 (Ar-C-C), 120.7 (Ar-C-C), 94.2 (-C≡C-), 89.4 (-C≡C-), 30.5 (CH$_3$) ppm; **IR**: 2204 (-C≡C-), 1687 (-C=O) cm$^{-1}$; **MS** ES+: calcd. [M]+ 527.1134; found. 527.1114.

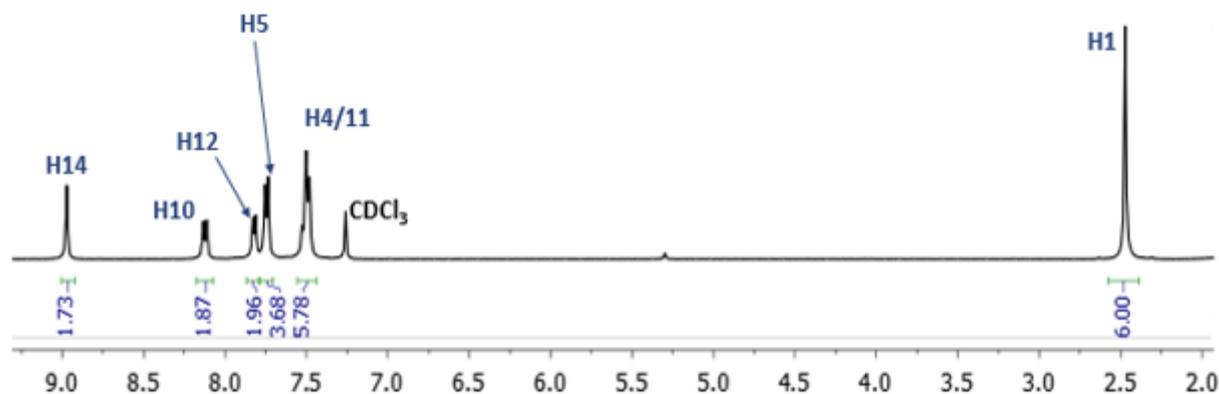

Figure S11: The $^1$H NMR spectrum of **4B** in CDCl$_3$

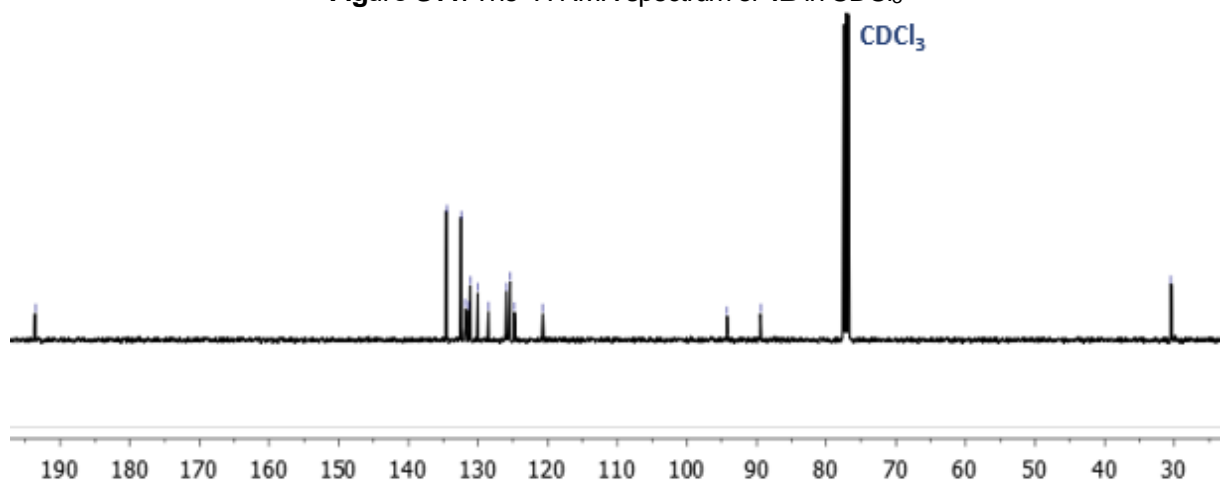

Figure S12: The $^{13}$C{$^1$H} NMR spectrum of **4B** in CDCl$_3$

### 1.4 Crystallographic Information

**The X-ray crystal structure of 1**

*Crystal data for* **1**: C$_{32}$H$_{22}$S$_2$, *M* = 470.61, monoclinic, *P*2$_1$/*c* (no. 14), *a* = 5.2790(5), *b* = 8.0061(9), *c* = 28.485(3) Å, β = 90.732(9)°, *V* = 1203.8(2) Å$^3$, *Z* = 2 [*C$_i$* symmetry], *D$_c$* = 1.298 g cm$^{-3}$, μ(Mo-Kα) = 0.240 mm$^{-1}$, *T* = 173 K, orange platy needles, Agilent Xcalibur 3 E diffractometer; 2437 independent measured reflections (*R*$_{int}$ = 0.0285), *F*$^2$ refinement,[6,7] *R$_1$*(obs) = 0.0538, *wR$_2$*(all) = 0.1280, 1795 independent observed absorption-corrected reflections [|*F*$_o$| > 4σ(|*F*$_o$|), completeness to θ$_{full}$(25.2°) = 99.8%], 156 parameters. CCDC 1944032.

The structure of **1** was found to sit across a centre of symmetry at the middle of the anthracenyl moiety.

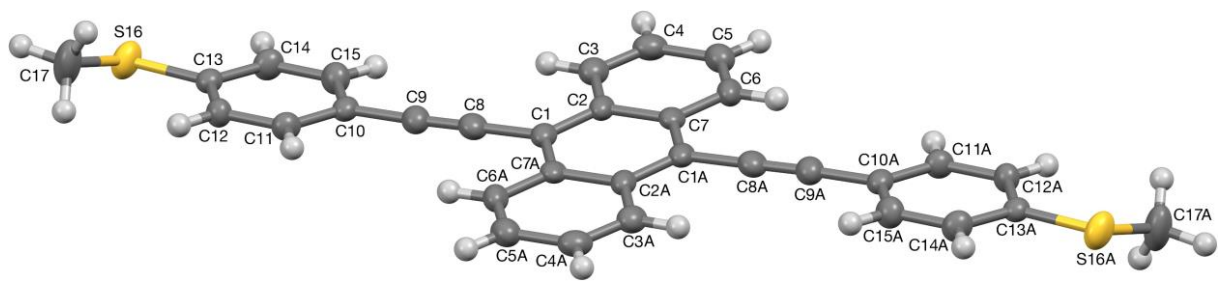

**Fig. S13** The crystal structure of the $C_i$-symmetric molecule **1** (50% probability ellipsoids).

### The X-ray crystal structure of 3

*Crystal data for* **3**: $C_{34}H_{22}O_2S_2$, $M$ = 526.63, triclinic, $P$-1 (no. 2), $a$ = 5.0612(10), $b$ = 9.591(2), $c$ = 13.810(2) Å, α = 98.634(15), β = 90.898(14), γ = 101.338(18)°, $V$ = 649.1(2) Å$^3$, $Z$ = 1 [$C_i$ symmetry], $D_c$ = 1.347 g cm$^{-3}$, μ(Mo-Kα) = 0.236 mm$^{-1}$, $T$ = 173 K, orange blocky needles, Agilent Xcalibur 3 E diffractometer; 2554 independent measured reflections ($R_{int}$ = 0.0229), $F^2$ refinement,[6,7] $R_1$(obs) = 0.0437, $wR_2$(all) = 0.0993, 1923 independent observed absorption-corrected reflections [|$F_o$| > 4σ(|$F_o$|), completeness to θ$_{full}$(25.2°) = 98.7%], 174 parameters. CCDC 1958589.

The structure of **3** was found to sit across a centre of symmetry at the middle of the anthracenyl moiety.

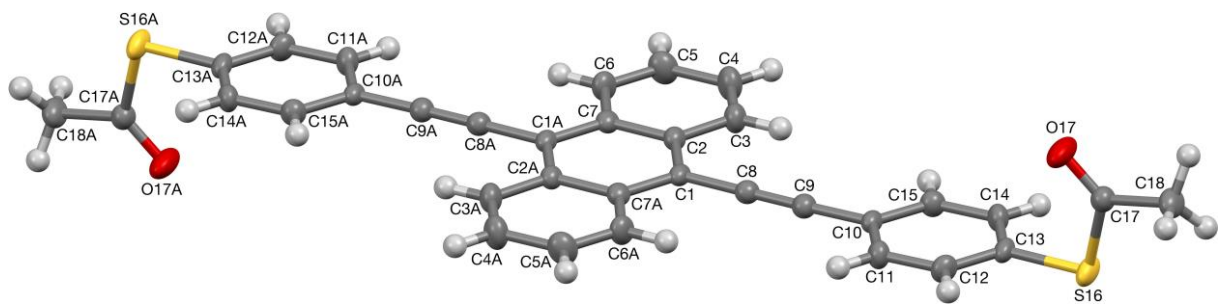

**Fig. S14** The crystal structure of the $C_i$-symmetric molecule **3** (50% probability ellipsoids).

### The X-ray crystal structure of 3A

*Crystal data for* **3A**: $C_{38}H_{34}S_2$, $M$ = 554.77, monoclinic, $C2$ (no. 5), $a$ = 27.0814(8), $b$ = 5.9789(2), $c$ = 18.7842(6) Å, β = 97.344(3)°, $V$ = 3016.54(18) Å$^3$, $Z$ = 4, $D_c$ = 1.222 g cm$^{-3}$, μ(Cu-Kα) = 1.774 mm$^{-1}$, $T$ = 203 K, orange platy needles, Agilent Xcalibur PX Ultra A diffractometer; 4200 independent measured reflections ($R_{int}$ = 0.0221), $F^2$ refinement,[6,7] $R_1$(obs) = 0.0414, $wR_2$(all) = 0.1063, 3416 independent observed absorption-corrected reflections [|$F_o$| > 4σ(|$F_o$|), completeness to θ$_{full}$(67.7°) = 97.6%], 368 parameters. The structure of **3A** was refined as a two-component inversion twin [Flack parameter $x$ = 0.46(2)]. CCDC 1958590.

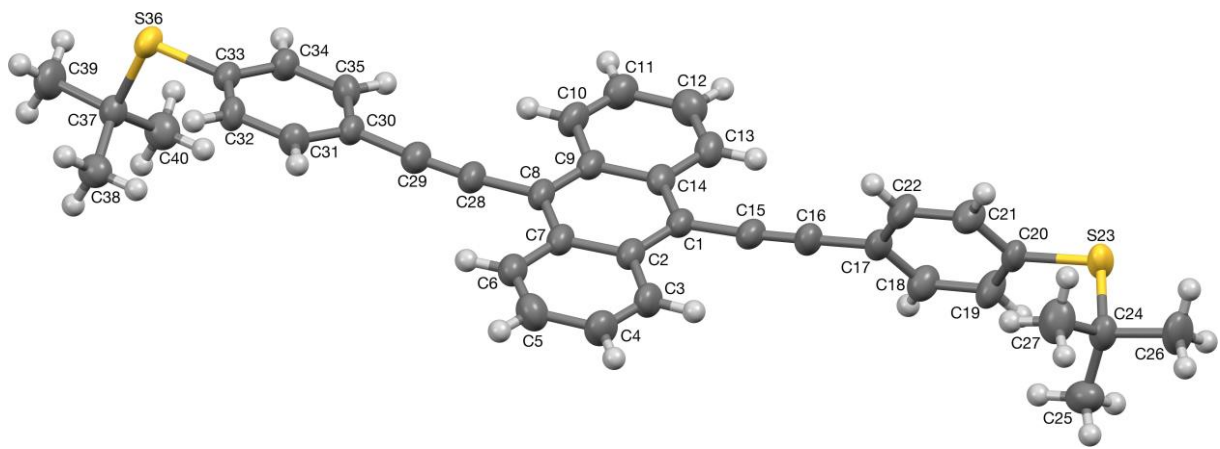

**Fig. S15** The crystal structure of **3A** (50% probability ellipsoids).

**The X-ray crystal structure of 4A**

*Crystal data for* **4A**: $C_{38}H_{34}S_2$, $M$ = 554.77, triclinic, $P$-1 (no. 2), $a$ = 5.9983(6), $b$ = 10.9346(10), $c$ = 11.9346(11) Å, α = 95.669(7), β = 97.009(8), γ = 93.405(8)°, $V$ = 771.11(13) Å$^3$, $Z$ = 1 [$C_i$ symmetry], $D_c$ = 1.195 g cm$^{-3}$, μ(Cu-Kα) = 1.735 mm$^{-1}$, $T$ = 203 K, yellow plates, Agilent Xcalibur PX Ultra A diffractometer; 2914 independent measured reflections ($R_{int}$ = 0.0370), $F^2$ refinement,[6,7] $R_1$(obs) = 0.0437, $wR_2$(all) = 0.1226, 2182 independent observed absorption-corrected reflections [|$F_o$| > 4σ(|$F_o$|), completeness to θ$_{full}$(67.7°) = 98.0%], 184 parameters. CCDC 1958591.

The structure of **4A** was found to sit across a centre of symmetry at the middle of the anthracenyl moiety.

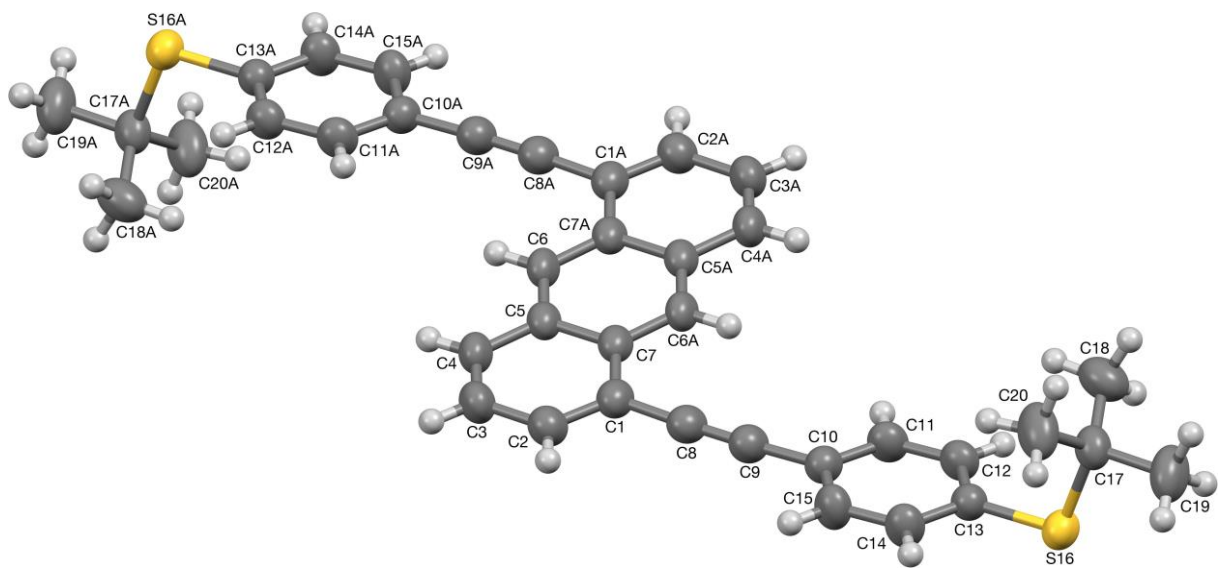

**Fig. S16** The crystal structure of the $C_i$-symmetric molecule **4A** (50% probability ellipsoids)

## 2. Additional DFT and Transport Calculations

### 2.1 Optimised DFT Structures of Isolated Molecules

Using the density functional code SIESTA,[8,9] the optimum geometries of the isolated molecules **1-4** were obtained by relaxing the molecules until all forces on the atoms were less than 0.01 eV / Å as shown in Figure S17. A double-zeta plus polarization orbital basis set, norm-conserving pseudopotentials, an energy cut-off of 250 Rydbergs defined the real space grid were used and the local density approximation (LDA) was chosen to be the exchange correlation functional. We also computed results using GGA and found that the resulting transmission functions were comparable with those obtained using LDA.[10,11]

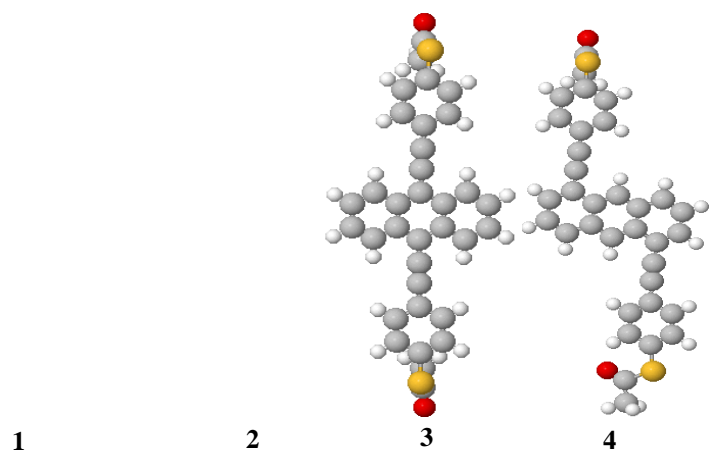

**1**   **2**   **3**   **4**

**Figure S17**: Fully relaxed isolated molecules. Key: C = grey, H = white, O = red, S = yellow.

### 2.2 Frontier orbitals of the molecules

The plots below show isosurfaces of the HOMO, LUMO, HOMO-1 and LUMO+1 of isolated molecules **1-4**.

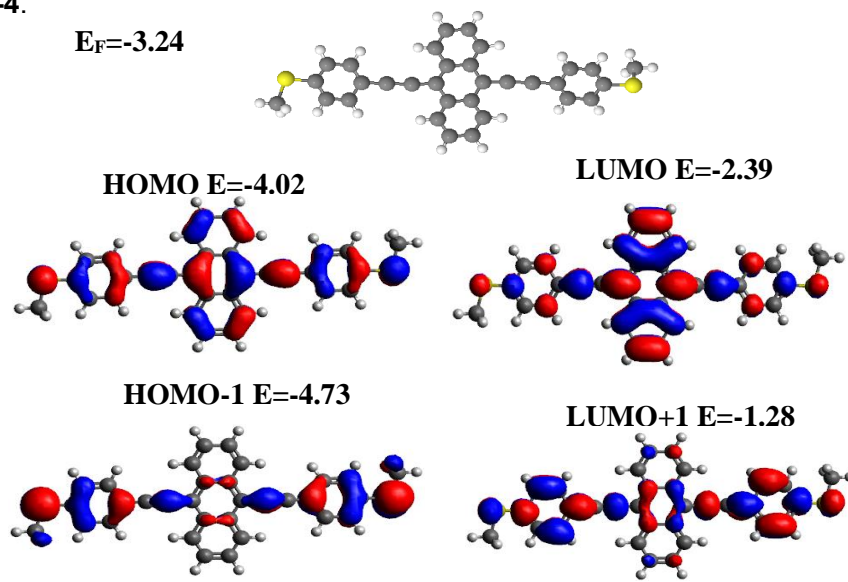

$E_F = -3.24$

HOMO E=-4.02   LUMO E=-2.39

HOMO-1 E=-4.73   LUMO+1 E=-1.28

**Figure S18**: Wave function for **1**. Top panel: Fully optimsed geometry of **1**. Lower panel: HOMO, LUMO, HOMO-1 and LUMO+1 along with their energies

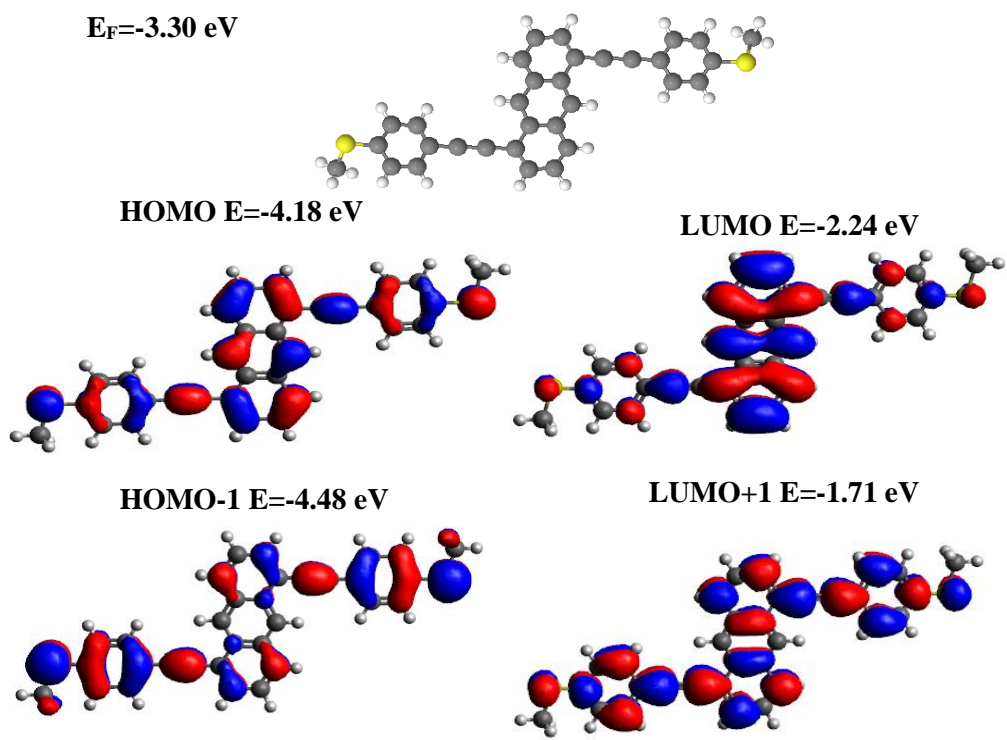

**Figure S19**: Wave function for **2**. Top panel: Fully optimised geometry of **2**. Lower panel: HOMO, LUMO, HOMO-1 and LUMO+1 along with their energies

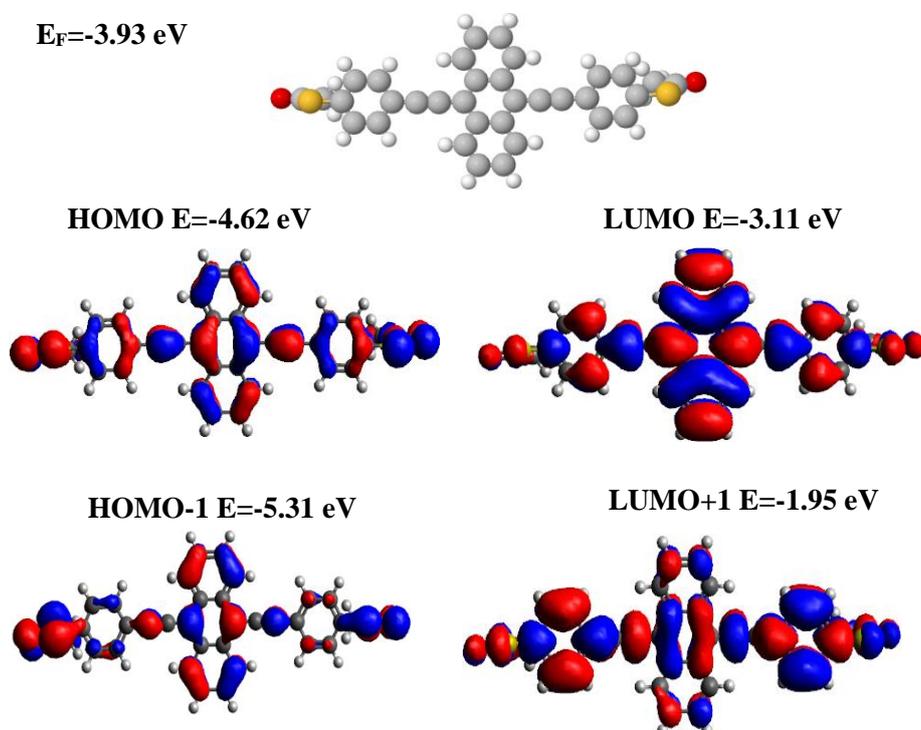

**Figure S20**: Wave function for **3**. Top panel: Fully optimised geometry of **3**. Lower panel: HOMO, LUMO, HOMO-1 and LUMO+1 along with their energies

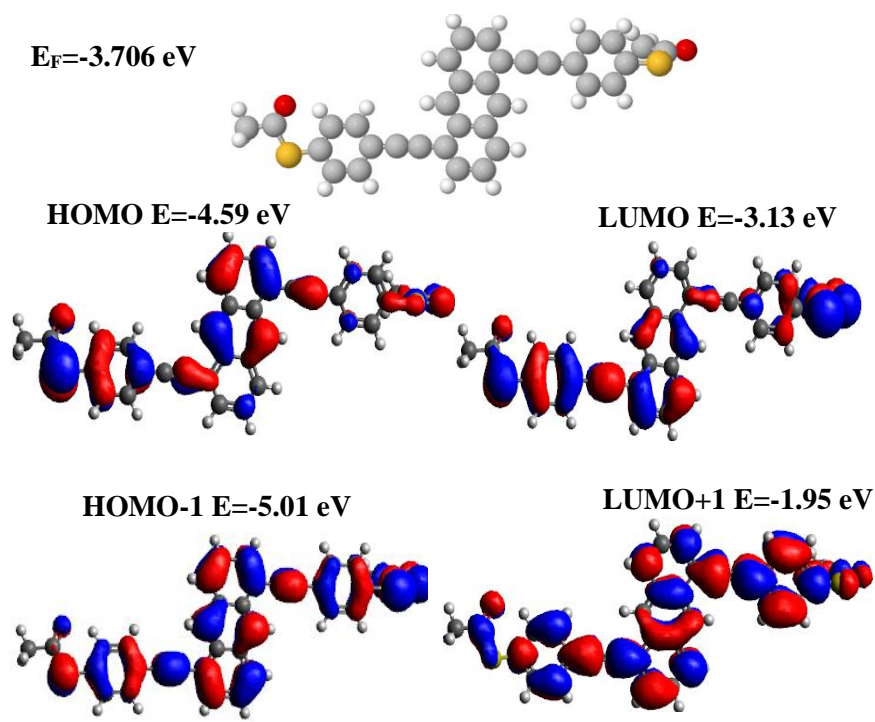

**Figure S21**: Wave function for **4**. Top panel: Fully optimised geometry of **4**. Lower panel: HOMO, LUMO, HOMO-1 and LUMO+1 along with their energies.

## 2.3 Product rule

Wave function plots for isolated molecules with their optimised geometries (Figures S14-S17) show iso-surfaces of the HOMO, LUMO, HOMO-1 and LUMO+1 of isolated molecules of the studied molecules. The information of the product rule[12-14] is obtained from Figures S14-S17. Product rule predicts a CQI in the HOMO-LUMO gap for the molecules of study, because the product of the HOMO (LUMO) amplitudes at opposite ends of the molecules is negative (positive). Table S1 summarises the signs of these orbital products.

**Table S1**: Product rule predictions of the studied molecules, (c= constructive, d=destructive, blue= -ive and red= +ive).

| Compound | H-1   | H     | L     | L+1   | $G_{H\text{-}L}$ |
|----------|-------|-------|-------|-------|------------------|
| **1**    | +     | -     | +     | -     | c                |
| E (eV)   | -4.73 | -4.02 | -2.39 | -1.28 |                  |
| **2**    | +     | -     | +     | -     | c                |
| E (eV)   | -4.48 | -4.18 | -2.24 | -1.71 |                  |
| **3**    | +     | -     | +     | -     | c                |
| E (eV)   | -5.31 | -4.62 | -3.11 | -1.95 |                  |
| **4**    | -     | -     | +     | -     | c                |
| E (eV)   | -5.01 | -4.59 | -3.13 | -1.95 |                  |

## 2.4 Binding energy of molecules on Au

To calculate the optimum binding distance between pyridyl/methyl sulphide anchor groups and Au(111) surfaces, we used DFT and the counterpoise method, which removes basis set superposition errors (BSSE). The binding distance d is defined as the distance between the gold surface and the S/SMe terminus of the thiol/methyl sulphide group. Here, compound **1** is defined as entity A and the gold electrode as entity B. The ground state energy of the total system is calculated using SIESTA and is denoted $E_{AB}^{AB}$. The energy of each entity is then calculated in a fixed basis, which is achieved using ghost atoms in SIESTA. Hence, the energy of the individual **1** in the presence of the fixed basis is defined as $E_A^{AB}$ and for the gold as $E_B^{AB}$. The binding energy is then calculated using the following equation:

$$\text{Binding Energy} = E_{AB}^{AB} - E_A^{AB} - E_B^{AB} \quad\quad\quad (S1)$$

We then considered the nature of the binding depending on the gold surface structure. We calculated the binding to a Au pyramid on a surface with the thiol/methyl sulphide atom binding at a 'top' site and then varied the binding distance d. Figure S22 (left) shows that a value of d = 2.4 Å gives the optimum distance, at approximately 0.8 eV. As expected, the thiol anchor group binds favourably to under-coordinated gold atoms. For SMe d = 2.7 Å gives the optimum distance, at approximately 0.5 eV.

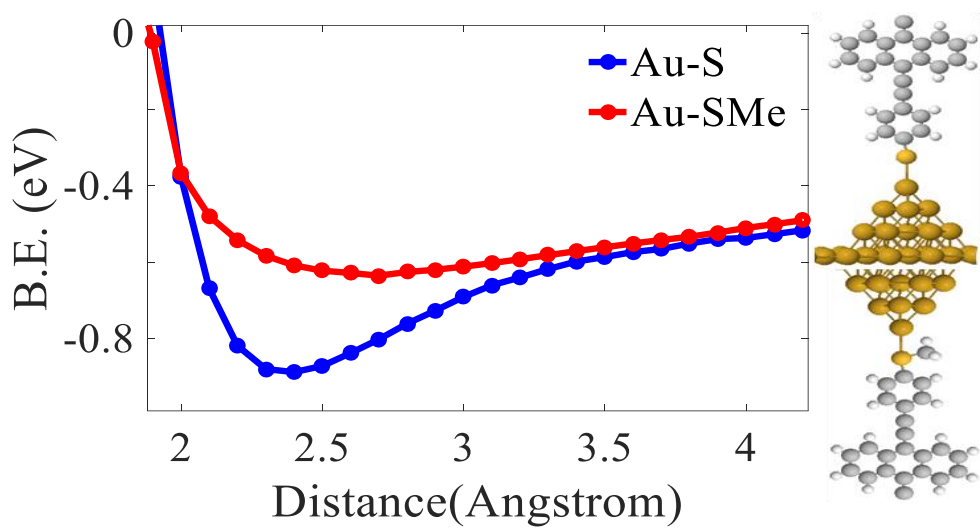

**Figure S22**: Example binding energy plot of **4**, for two different anchors Au-S and Au-SMe (left), with its idealised ad-atom configuration at the Au lead interface (right, top Au-S and bottom Au-SMe).
Key: C = grey, H = white, S = light yellow, Au = dark yellow.

**2.5 Optimised DFT Structures of Compounds in Their Junctions**

Using the optimised structures and geometries for the compounds obtained as described in section 2.1 (above), we again employed the SIESTA code to calculate self-consistent optimised geometries, ground state Hamiltonians and overlap matrix elements for each metal-molecule-metal junction. Leads were modelled as 625 atom slabs, terminated with 11-atom Au(111) tips. The optimised structures were then used to compute the transmission curve for each compound. The DFT optimised geometries are shown here, in Figures S23-26. Note: there is a tilt angle range for each compound, which presents in section 2.5.

Key: C = grey, H = white, S = light yellow, Au = dark yellow.

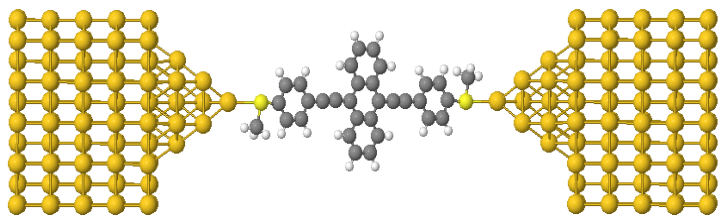

**Figure S23**: Optimised structure of **1**.

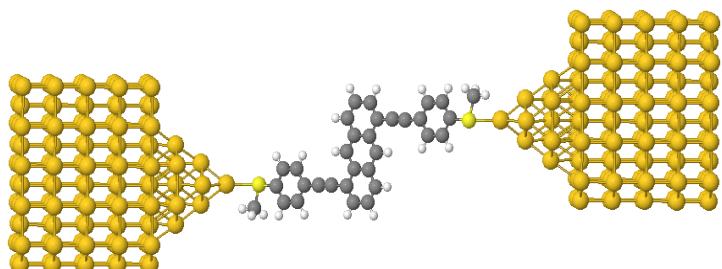

**Figure S24**: Optimised structure of **2**.

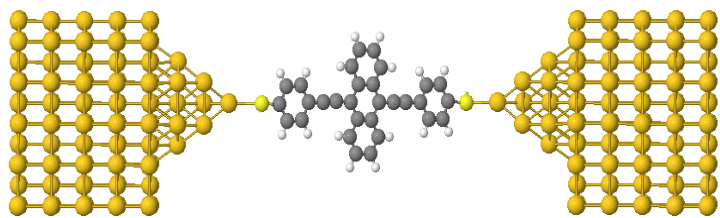

**Figure S25**: Optimised structure of **3**.

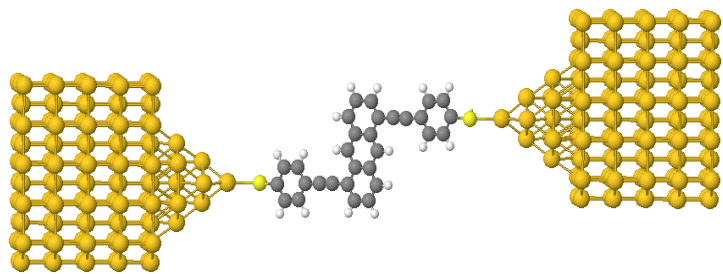

**Figure S26**: Optimised structure of **4**.

## 2.6 The tilt angle (θ)

In this section, we determine the tilt angle $\theta$ of each compound on a gold substrate, which corresponds to the experimentally measured most-probable break-off distance. In previous work [15] we have demonstrated how the tilt angle calculates for both single molecule and SAM. Table S2 shows each compound for a range of tilt angles. Break-off distance values suggest that compound-**1** and -**2** tilt with angle θ ranging from 51º to 59º, compound-3 29º-33º and compound-4 27º-35º

**Table S2:** Experimental break-off distance and equivalent tilt angle (θ)

| Compound | Experimental film thickness (nm) | Experimental film roughness (nm) | Equivalent experimental tilt angle (θ) | Equivalent theoretical tilt angle (θ) |
|---|---|---|---|---|
| 1 | 1.25 | 0.17 | 51°-59° | 51°-59° |
| 2 | 1.28 | 0.11 | 51°-59° | 51°-59° |
| 3 | 1.12 | 0.43 | 29°-33° | 29°-33° |
| 4 | 1.19 | 0.09 | 27°-35° | 27°-35° |

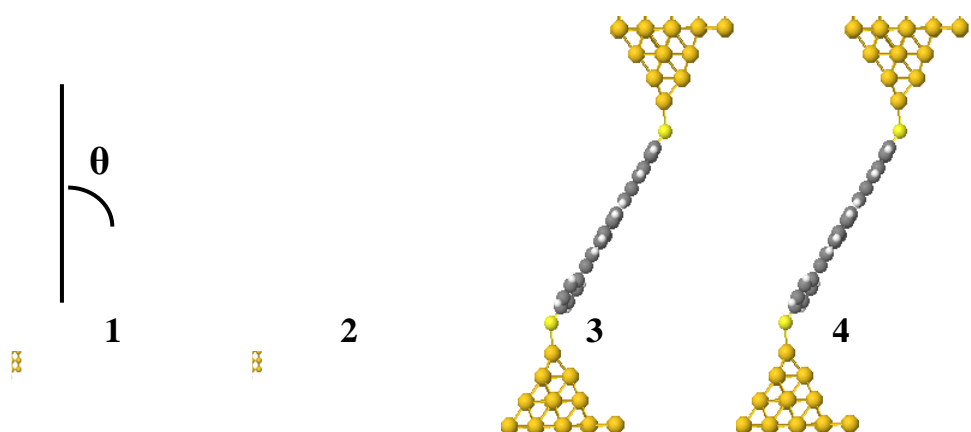

**Figure S27**: Optimised structures of **1-4**. Tilt angle (side-view)

## 2.7 HOMO-LUMO gaps

The calculated and optically measured HOMO-LUMO gaps are listed in Table S3. Theoretical gaps were calculated for isolated molecules and when the compounds are placed in the junctions, the gap between their HOMO and LUMO transmission resonances are quoted. As shown by the third and fourth columns in Table S2, isolated gaps for compounds **1**, **2**, **3** and **4** are larger than the gaps between the transmission resonances. This is because the latter are shifted by the real part of the self-energy of the contact to the leads, reflecting the fact that the system is more open when contacted to electrodes. In general, theoretical gaps are smaller than the measured gaps, which is consistent with the fact that DFT is known to underestimate its value.[16,17]

**Table S3**: Experimental and theoretical HOMO–LUMO gaps in eV.

| Compound | $E_g$, (Exp.) a | $E_g$, DFT (Iso.) b | $E_g$, DFT (Au-M-Au) |
|---|---|---|---|
| 1 | 2.57 | 1.63 | 1.30 |
| 2 | 3.75 | 1.94 | 1.85 |
| 3 | 2.66 | 1.51 | 1.20 |
| 4 | 2.61 | 1.46 | 1.25 |

[a] Experimental data: $E_g = 1241.5/\lambda_{ABS}$. [b] Theoretical HOMO–LUMO gaps for the isolated molecules. [c] Theoretical gaps between HOMO–LUMO transmission resonances in Au|molecule|Au structures.

## 2.8 Transport Calculations

The transmission coefficient curves $T(E)$, obtained from using the Gollum transport code, were calculated for compounds **1-4** based on the tilt angle range in Table S1. The LUMO resonance is predicted to be pinned near the Fermi Level of the electrodes for the four molecules, however, we choose Fermi Level to be in the mid gap at approximately ±0.5 eV (black-dashed line), as shown in Figure S28.

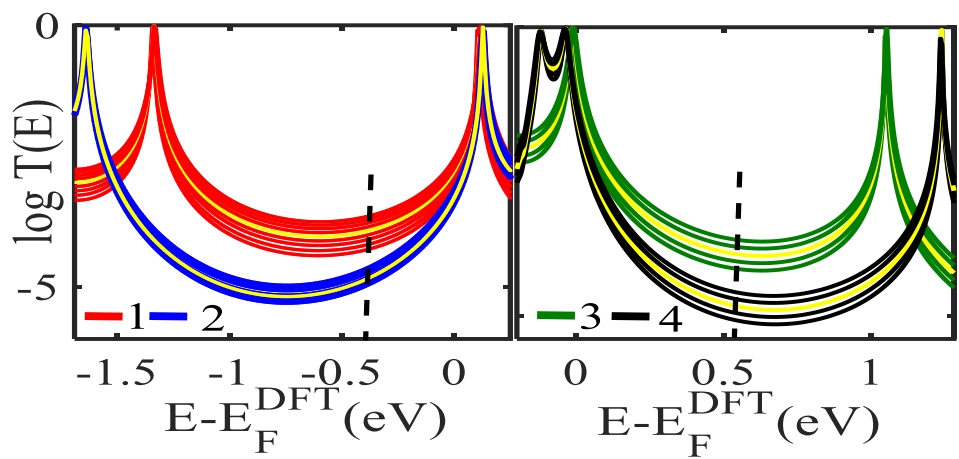

**Figure S28**: Zero bias transmission coefficient *T(E)* of molecules **1-4** against electron energy E. **Left panel:** compound **1** (red-lines), compound **2** (blue-line). **Right panel:** compound **3** (green-line), and compound **4** (black-line), yellow-lines the average of each compound.

In previous works[15] we have demonstrated that the transmission coefficient T(E), for single molecule is approximately the same for SAM, by comparing T(E) for single molecule against SAM consists of 7 molecules.

## 2.9 Seebeck coefficient

After covering the electronic transport for the four molecules, the study of some thermoelectronic properties such as thermopower $S$ for the same groups is made.

To calculate the thermopower of these molecular junctions, it is useful to introduce the non-normalised probability distribution $P(E)$ defined by

$$P(E) = -\mathcal{T}(E)\frac{df(E)}{dE} \qquad (S2)$$

where $f(E)$ is the Fermi-Dirac function and $\mathcal{T}(E)$ are the transmission coefficients and whose moments $L_i$ are denoted as follows

$$L_i = \int dE P(E)(E - E_F)^i \qquad (S3)$$

where $E_F$ is the Fermi energy. The thermopower, $S$, is then given by

$$S(T) = -\frac{1}{|e|T}\frac{L_1}{L_0} \qquad (S4)$$

where *e* is the electronic charge.

Supplementary Figure S29 shows the thermopower $S$ evaluated at room temperature for different energy range $E_F - E_F^{DFT}$.

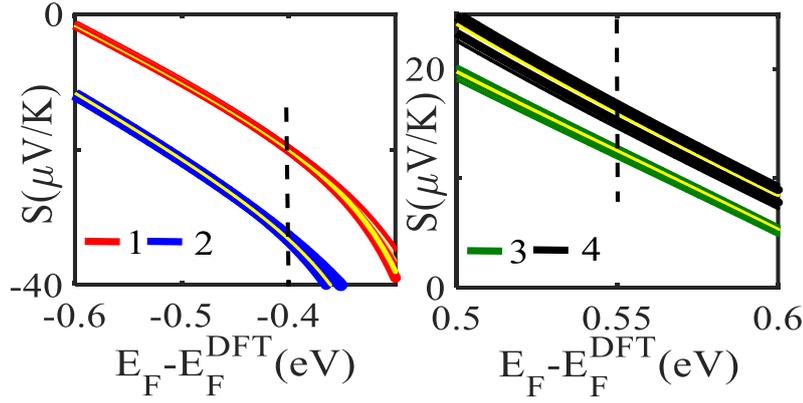

**Figure S29**: Seebeck coefficient S of the same molecules. **Left panel**: (**1** and **2**), **Right panel**: (**3** and **4**)

## 3. Experimental part

### 3.1 QCM monitoring of SAMs growth and single molecular occupation area estimation

The difference between the measured frequency and initial frequency, $\Delta f$, related with the amount of molecule on Au surface is given by the Sauerbrey equation:

$$n = \frac{-\Delta f \times A \times k \times N_A}{M_w} \tag{S5}$$

$$k = \frac{\sqrt{\mu * \rho}}{2 * f_0^2} \tag{S6}$$

Where *n* the amount of molecule adsorbed on Au surface, *A* the electrode area, $N_A$ the Avogadro's number, $M_w$ the molecular weight, *μ* the shear modulus of quartz, *ρ* the density of quartz, *f0* the initial frequency.

The single molecular occupation area on Au surface can be calculated by:

$$A_{molecule} = \frac{A_{Electrode}}{n} \tag{S7}$$

the corresponding results were listed in Table s3.

### 3.2 Single molecular conductance calculation

The plot of dI/dV (S) vs. bias voltage was shown in Figure s26. The number of molecules contacted by the probe was calculated using contact area between sample and probe dividing the occupation area of a single molecule. The contact area between sample and probe was estimated by Hertzian model:

$$r = (F \times R \times \frac{1}{Y})^{\frac{1}{3}}$$

$$\frac{1}{Y} = \frac{3}{4} \times (\frac{1-v_1^2}{E_1} + \frac{1-v_2^2}{E_2})$$

Where *r* the contact radius, *F* the loading force from probe to sample, *R* the radius of the probe (~10 nm from the supplier), $v_1$ and $v_2$ the Poisson ratio of the material, $E_1$ and $E_2$ the Young's

Modulus for probe (~ 100 GPa) and SAMs (~10 GPa, estimated by nano-mechanical mapping under peak force mode).

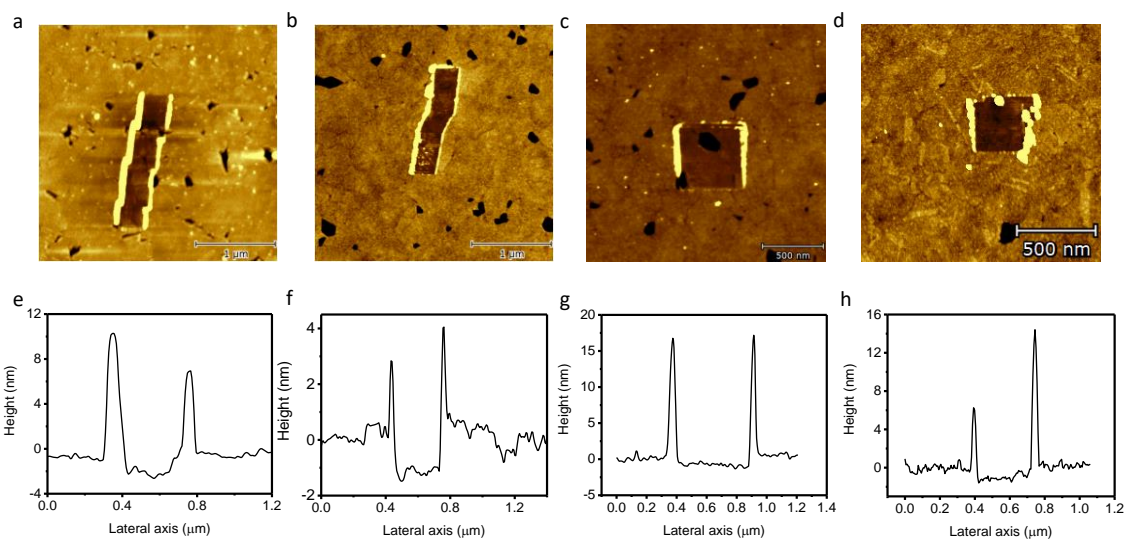

**Figure S30:** (a-d) AFM topography of molecules **1-4** SAMs after nano-scratching by AFM probe. (e-h) Height profile for (a-d) correspondingly.

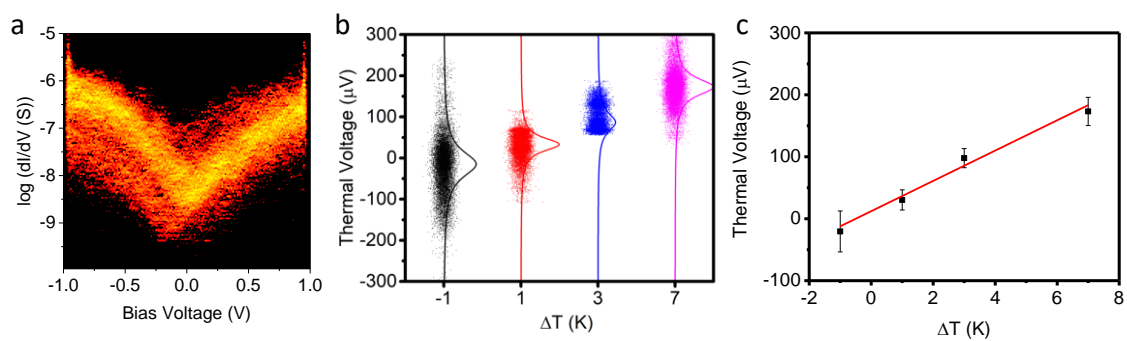

**Figure S31:** Electric and thermoelectric properties of molecule 1. (a) Heat map of electrical conductance vs. applied bias voltage by cAFM. (b) Measured thermal voltage vs. ΔT ($T_{sample}$ - $T_{probe}$). (c) Linear curve fit of thermal voltage vs. ΔT.

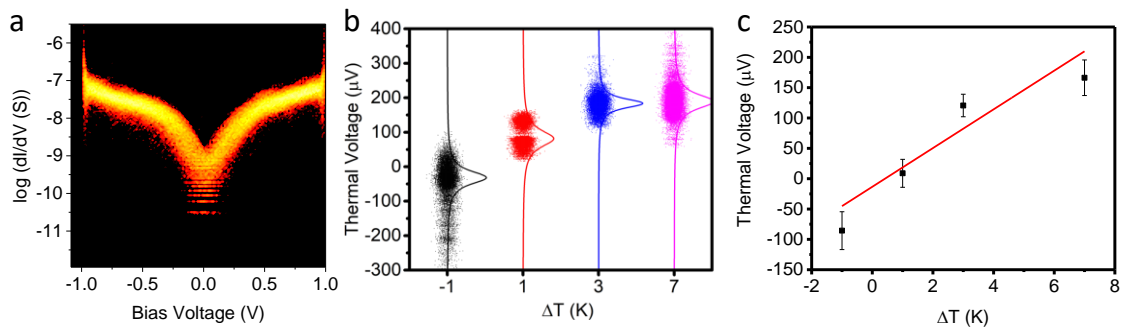

**Figure S32:** Electric and thermoelectric properties of molecule 2. (a) Heat map of electrical conductance vs. applied bias voltage by cAFM. (b) Measured thermal voltage vs. ΔT ($T_{sample}$ - $T_{probe}$). (c) Linear curve fit of thermal voltage vs. ΔT.

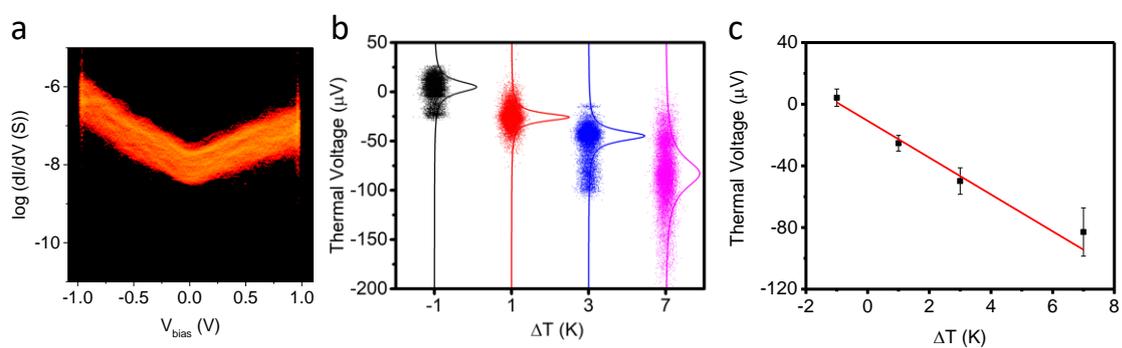

**Figure S33:** Electric and thermoelectric properties of molecule 3. (a) Heat map of electrical conductance vs. applied bias voltage by cAFM. (b) Measured thermal voltage vs. ΔT ($T_{sample}$ - $T_{probe}$). (c) Linear curve fit of thermal voltage vs. ΔT.

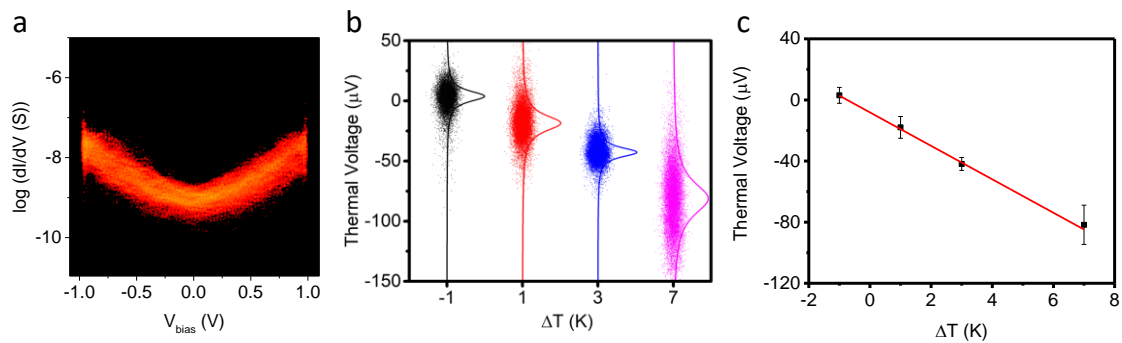

**Figure S34:** Electric and thermoelectric properties of molecule 4. (a) Heat map of electrical conductance vs. applied bias voltage by cAFM. (b) Measured thermal voltage vs. ΔT ($T_{sample}$ - $T_{probe}$). (c) Linear curve fit of thermal voltage vs. ΔT.

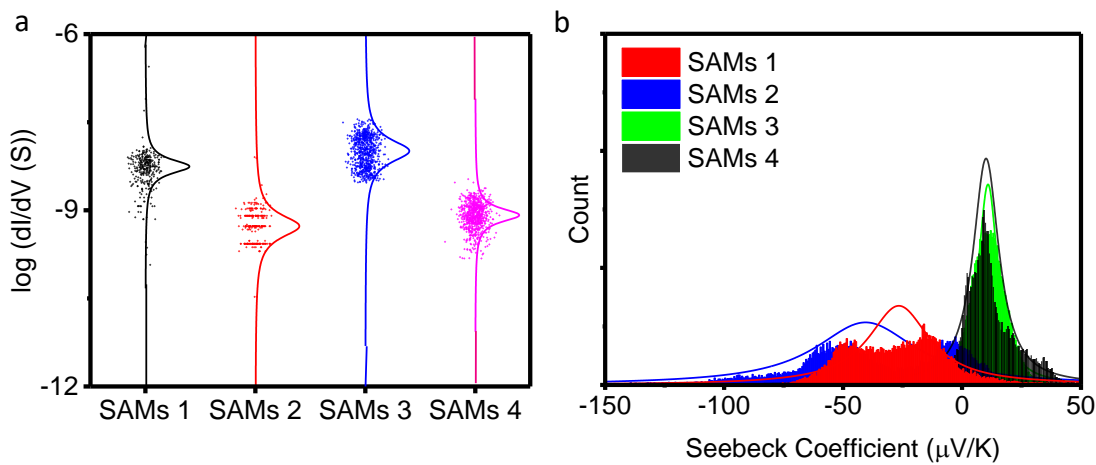

**Figure S35（a）** Electrical conductance distribution of SAMs 1-4 at low bias voltage (-20 mV to 20 mV) (b) Histogram of Seebeck coefficient distribution of SAMs 1-4.

**Table S4:** Single molecule occupation area from QCM

|  | Δf (Hz) | m (ng) | Mw (g/mol) | A (Ǎ$^2$) |
|---|---|---|---|---|
| Molecule **1** | 104 | 90.7 | 471 | 33.9 |
| Molecule **2** | 115 | 100.2 | 471 | 30.6 |
| Molecule **3** | 99 | 86.3 | 444[1]/487[2] | 34.2[1]/37.8[2] |
| Molecule **4** | 93 | 99.4 | 444/487 | 35.8/39.2 |

[1] if SAMs terminated with SH

[2] if SAMs terminated with SAc

\* A is the occupation area of a single molecule